\documentclass[11pt]{article}

\usepackage{geometry,graphicx,amsmath,amssymb,amsfonts,mathdots,mathrsfs,caption,subcaption, setspace, wrapfig}
\usepackage[usenames,dvipsnames,svgnames,table]{xcolor}
\pdfoutput=1
\usepackage{graphicx}
\usepackage{jheppub}
\usepackage{epsfig}
\usepackage{epstopdf}
\usepackage{longtable}
\usepackage{braket}
\usepackage{wasysym}
\usepackage{comment}
\usepackage{amsmath}
\usepackage{slashed}
\usepackage{tabularx}
\usepackage{multicol}
\usepackage{multirow}
\usepackage{xspace}
\usepackage{xcolor}
\usepackage{comment}
\usepackage{enumitem}

\graphicspath{{../}{./Results/}}
\newcommand{\met}{\ensuremath{\not\!\!E_T}\xspace}

\usepackage{listings}
\usepackage{color}

\definecolor{dkgreen}{rgb}{0,0.6,0}
\definecolor{gray}{rgb}{0.5,0.5,0.5}
\definecolor{mauve}{rgb}{0.58,0,0.82}

\newcommand*{\file}{\fontfamily{phv}\selectfont}

\begin{document}


\title{Seer: An analysis package for LHCO files.}
\author[]{Travis A. Martin}
\affiliation[]{TRIUMF, 4004 Wesbrook Mall, Vancouver, Canada V6T 2A3}
\emailAdd{tmartin@triumf.ca}

\abstract{Seer is a multipurpose package for performing trigger, signal determination and cuts of an arbitrary number of collider processes stored in the LHCO file format. This article details the use of Seer, including the necessary details for users to customize the code for investigating new kinematic variables.\\
Seer can be downloaded at \url{http://sourceforge.net/p/seerhep/}}

\maketitle

\section{Introduction\label{sec:intro}}

Many code packages have been created to perform rapid Monte Carlo integration of the phase space for collider physics processes, to simulate the effects of hadronization, and to emulate the response of detectors such as ATLAS and CMS. The most commonly used event generator is arguably MadGraph5+aMC@NLO \cite{Alwall:2014hca}, with Pythia \cite{Sjostrand:2007gs}, Herwig++ \cite{Bahr:2008pv}, Sherpa \cite{Gleisberg:2003xi} and WHiZaRD \cite{Reuter:2014ema} also commonly used for event generation. To simulate the detector response, including efficiency and isolation details, PGS \cite{Carena:2000yx} and Delphes \cite{Selvaggi:2014mya} are the two most commonly used packages for theorists/phenomenologists, with a modified version of Delphes being employed by the CheckMATE \cite{Drees:2013wra} code package, while full detector simulators are available only to experimental groups. The collection and analysis of the events passing through the generation-through-detector-simulation process is often left to the individual, with some packages, such as MadAnalysis \cite{Conte:2014xya}, existing to assist in the reproduction of experimental analyses, with varying learning curves.

These tools are important for phenomenologists for two purposes -- proposing new analyses that will be sensitive to a well motivated model or signature, and reproducing experimental analyses for a particularly motivated model to determine exclusions of parameter space. One of the challenges of performing these studies is in producing a sufficiently large number of events, especially of standard model (SM) backgrounds, to achieve an appropriate coverage of the full phase space where the analysis focuses. A good rule-of-thumb in generating backgrounds to reproduce an experimental analysis is to generate a sufficiently large number of events that the per-event weight is equal to the inverse of the integrated luminosity. Thus, any process with a nanobarn cross section at LHC8 with 20/fb integrated luminosity, would require 20 million events to be generated. Techniques can be employed at the event generation stage to focus on the region of phase space of the SM backgrounds which overlap with the signal of interest, however this limits the reusability of these backgrounds for scans over parameter space that result in varying kinematic signatures. As a result, many analyses would require a large number of events to be generated.

Common output file formats from the generation--hadronization--simulation process, including the Les Houches file format \cite{Alwall:2006yp} and the ROOT file format \cite{Antcheva:2009zz}, produce large file sizes that can be prohibitive to store for researchers without significant computing resources. Alternatively, the LHC Olympics (LHCO) file format is a minimalist output file format that can be produced by both PGS and Delphes detector simulation packages, and has the added bonus of shielding the user from all truth information regarding the events, thus eliminating the possibility of user bias in an analysis. This file format allows a user to store a large number of events with relatively little disk space use.

This article introduces a new analysis package, entitled Seer, for the rapid analysis of an arbitrary number of LHCO files (limited by computer memory only, tested with over 850 files of 50k events each, simultaneously).\footnote{A similar effort has been made with the CutLHCO program \cite{Walker:2012vf}. In addition, MadAnalysis can perform many of the same functions, though Seer offers some differences in both the input and output, and provides some different features.} The goal of this package is to provide a simple way to combine multiple event generation files for many different processes, apply trigger requirements and kinematic cuts, and analyze and histogram the results. Many common kinematic variables are already implemented, including the popular stransverse mass ($M_{T2}$) \cite{Cheng:2008hk,Lester:1999tx,Barr:2003rg} variable and the Razor variables\footnote{Razor calculations use the script provided by CMS at \url{https://twiki.cern.ch/twiki/bin/view/CMSPublic/RazorLikelihoodHowTo}} \cite{Rogan:2010kb,Chatrchyan:2012uea} used to analyze SUSY decays, allowing usefulness out-of-the-box. Since the package employs ROOT libraries, modification of the code to add user defined kinematic calculations and cuts is simple and easy for new users.

The installation of Seer is covered in Section \ref{sec:install}. Following this, an in-depth discussion regarding the use of Seer is covered in Section \ref{sec:use}, focusing on the ``out-of-the-box" version. Instructions for modifying Seer to creatie user-defined cuts and kinematic variables is discussed in Section \ref{sec:mod}. Lastly, a series of examples are provided showing very basic analyses of Seer, and examples of the final output histograms.

Of note, a variety of font face formats have been adopted to differentiate between references to files, references to text settings, and references to parts of the Seer code. As a brief legend, files use a {\file helvetica font}, text settings use a \texttt{true-type font}, and references to code use a {\small\texttt{small true-type font}}.

\section{Installing Seer\label{sec:install}}

Installing Seer is straightforward. Seer has been tested on OS X 10.7+ and Linux distributions using the GNU C++ compilers. Future versions may address other compiler options.

\subsection{Pre-requisites}

Seer requires ROOT libraries, as much of the functionality uses classes defined within ROOT. As a result, ROOT needs to be installed. Instructions regarding the installation of ROOT can be found at \url{https://root.cern.ch/drupal/content/downloading-root}. Seer requires that the \texttt{\$ROOTSYS} system variable be correctly setup in order to function.

\subsection{Co-requisites}

Seer includes a copy of the MT2/stransverse mass calculator developed by Z. Han and H.C. Cheng \cite{Cheng:2008hk,Lester:1999tx,Barr:2003rg}. To implement alternative versions of this code, the {\file mt2\_bisect.cpp} and {\file mt2\_bisect.h} files should be replaced in the Seer main directory.

Any extra C++ code a user wishes to add can be placed in the Seer directory and added to the {\file makefile}. The name of the header file needs to be added to the \texttt{\$HEADERS} variable in the {\file makefile}, and the object file to create needs to be added to the \texttt{\$OBJECTS} variable, in addition to adding the appropriate include statement in the Seer code.

\subsection{Unpacking}

Seer must first be unpacked. This can be performed by navigating to the directory containing the tarball and running the command {\em tar -zxvf seer.tar.gz}. This will unpack all the files and directories needed by Seer. The following files are mandatory for Seer to compile and run:
\begin{itemize}
\item {\file seer.h} \& {\file seer.cpp} -- header and C++ file for the main script.
\item {\file eventweights.h} \& {\file eventweights.cpp} -- header and C++ file for the script that pre-scans the LHCO files for cross section and number of events.
\item {\file calculations.h} \& {\file calculations.cpp} -- header and C++ file for the cuts/trigger/analysis class {\small\texttt{event}}.
\item {\file plots.h} \& {\file plots.cpp} -- header and C++ file for the plotting class {\small\texttt{plot}}.
\item {\file style.h} -- style file containing a number of settings for the histograms.
\item {\file mt2\_bisect.h} \& {\file mt2\_bisect.cpp} -- header and C++ file for the MT2 calculator.
\item {\file seer\_cuts.txt} -- contains all the cut and trigger settings
\item {\file seer\_fakes.txt} -- contains all the fake rate settings
\item {\file seer\_files.txt} -- contains the list of LHCO files
\item {\file seer\_plots.txt} -- contains the plot settings
\item {\file seer\_settings.txt} -- contains all settings for the constants that can be used in the calculations
\end{itemize}

\subsection{Compiling}

Seer is compiled by navigating to the directory in a terminal window and entering the $make$ command.

\section{Using Seer\label{sec:use}}

\begin{figure*}[!t]
\centering
\mbox{\includegraphics[width=\textwidth]{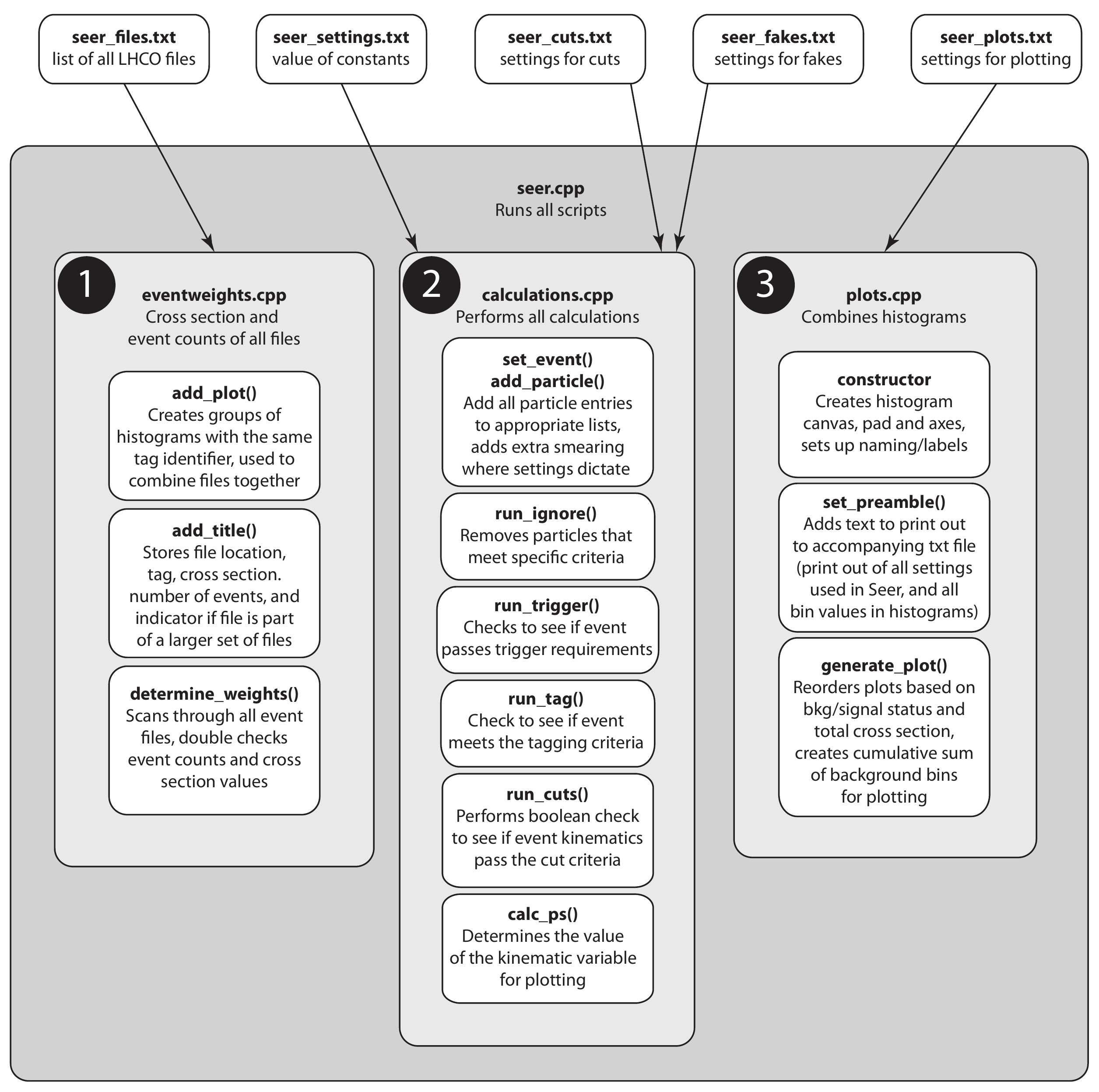}}
\caption{Flow chart of Seer, describing where input settings are used, and the main flow of processes, and the files that contain each of the main functions.
\label{fig:flowchart}}
\end{figure*}

Seer is run via a single command in a terminal, {\file ./seer.exe}. All details of the run are determined by the settings in five text files: three primary files ({\file seer\_files.txt}, {\file seer\_cuts.txt} and {\file seer\_plots.txt}) and two secondary files ({\file seer\_fakes.txt}, and {\file seer\_settings.txt}). The general process followed by Seer is given in Figure \ref{fig:flowchart}. A brief summary of the files is included below, and detailed information is provided in individual sections for each file.\\

\noindent{\file seer\_files.txt}

This is the first file that should be adjusted when starting an analysis. It contains the list of all LHCO files in the analysis, as well as details on their groupings. More information regarding this file is provided in Section \ref{sec:files}.\\

\noindent{\file seer\_cuts.txt}

This is the second file that should be adjusted. It contains the details of the tagging, the experimental trigger thresholds, fine tuning details of the detector limits, and details of the cuts. More information regarding this file is provided in Section \ref{sec:cuts}.\\

\noindent{\file seer\_plots.txt}

This is the last primary file to adjust. It contains the settings required to determine what type of histogram to make, as well as the details of the histogram and aspects of the output file names. More information regarding this file is provided in Section \ref{sec:plots}.\\

\noindent{\file seer\_fakes.txt}

This is a secondary file, as the settings within are useful only for advanced users. Seer can be used to handle fakes instead of the detector simulator used. More information regarding this file is provided in Section \ref{sec:fakes}.\\

\noindent{\file seer\_settings.txt}

Some additional settings are contained in this file, including numerical constants used by Seer to add additional smearing to jet and missing transverse energy kinematics. More information regarding this file is provided in Section \ref{sec:settings}.\\

\subsection{Seer Output}

The output of Seer is two files -- one plot file and one text file. The naming system for both of these files is identical except for the extension, and follows the scheme:
\begin{verbatim}
(plot_type)_(name_string)_n######.(file_type)
\end{verbatim}
where \texttt{(plot\_type)} is fixed in {\file plots.cpp} and depends only on the \texttt{(plotnum)} setting in {\file seer\_plots.txt}, \texttt{(name\_string)} is the setting in {\file seer\_plots.txt} that allows users to add a unique or descriptive text identifier to a particular analysis, and the sequence of numbers is the total sum of events of all files included in {\file seer\_files.txt}. The output file type for the histogram depends on the \texttt{(file\_type)} setting in {\file seer\_files.txt}, while the text output file is in a .txt format. More information regarding these settings is given in Section \ref{sec:plots}. Note that running Seer multiple types with the same \texttt{(plotnum)} and same files in {\file seer\_files.txt} will result in overwriting previous files. Users need to use the \texttt{(name\_string)} field to differentiate between applying different cuts or other analysis details that do not change the file name.

The plot output file contains a figure of the histograms of the various groupings of files included in {\file seer\_files.txt}. The text output file contains a header with a copy of every line in the five input settings files. Following that, it provides the following information:
\begin{enumerate}
\item Individual File Summaries: for each file, a section starting with the text ``Now running event file..." provides the file path and tag in a descriptive sentence, then gives a summary of the results of the analysis.
\begin{itemize}
\item File Tag -- tag given in {\file seer\_files.txt}
\item File Event Weight -- the event weight of the individual file (ignoring multiple files with same tag)
\item Combined Event Weight -- the event weight used when accounting for combined files ($w_f$)
\item Untagged Events -- the number of untagged events, and corresponding cross section ($\sigma_{\mathrm{untagged}} = n_{\mathrm{untagged}} \times w_f$)
\item Cut Events -- the number of events that were tagged but were cut in the analysis, and corresponding cross section ($\sigma_{\mathrm{cut}} = n_{\mathrm{cut}} \times w_f$)
\item Tagged \& Uncut Events -- the number of events that were tagged and passed the cuts of the analysis, and corresponding cross section ($\sigma_{\mathrm{included}} = n_{\mathrm{included}} \times w_f$)
\item Total Events -- the sum of the Untagged, Uncut and Tagged \& Uncut events, and corresponding cross section.
\item Cross Section -- the final, included cross section passed to the histogram, including associated theoretical error ($\delta \sigma_{\mathrm{included}} = \sigma_{\mathrm{included}}/\sqrt{n_{\mathrm{included}}}$; if no events pass cuts/tagging, $\delta\sigma_{\mathrm{included}} = w_f$ is used)
\end{itemize}
\item Group Summaries: includes all events from all files with the same tag in {\file seer\_files.txt}.
\begin{itemize}
\item Group Tag -- tag of the group of files, as defined in {\file seer\_files.txt}
\item Untagged Events -- the number of untagged events, and corresponding cross section ($\sigma_{\mathrm{untagged}} = n_{\mathrm{untagged}} \times w_f$)
\item Cut Events -- the number of events that were tagged but were cut in the analysis, and corresponding cross section ($\sigma_{\mathrm{cut}} = n_{\mathrm{cut}} \times w_f$)
\item Tagged \& Uncut Events -- the number of events that were tagged and passed the cuts of the analysis, and corresponding cross section ($\sigma_{\mathrm{included}} = n_{\mathrm{included}} \times w_f$)
\item Total Events -- the sum of the Untagged, Uncut and Tagged \& Uncut events, and corresponding cross section.
\item Cross Section -- the final, included cross section passed to the histogram, including associated theoretical error ($\delta \sigma_{\mathrm{group}} = \sqrt{\sum \delta\sigma_{\mathrm{individual}}^2}$)
\end{itemize}
\item Overall Summary: a list of the results from all files combined. This is the same information as in the group summaries, except now it includes all groups combined.
\item Breakdown of Cuts: a list of the cross section removed by each individual cut. Since multiple cuts can remove an event, these are not exclusive cross sections.
\item Number checks: a weight sum and total number of events check to make sure everything adds up correctly
\item Histogram results: a tab separated text list of the bin centre and cross section values for each histogram (in fb)
\end{enumerate}
Users who are interested in the total cross section before and after cuts can read this information directly from the relevant group summary in the text file. Users who are interested in distributions have a choice between the visual histogram and the text of the histogram values. Of note, the backgrounds in the text file are not summative, unlike the background histograms in the plot file, as discussed in Section \ref{sec:plots}.

The following sections describe in great detail the text files that Seer uses to read in all of the settings for the run. It is important to note that a line can be commented out by using double hashtags, \#\#.

\subsection{{\file seer\_files.txt} \label{sec:files}}

Seer checks this file for the list of all LHCO files to be used in the analysis. For each file/line, there are four required entries and two optional entries. The format of these entries is as follows:
\begin{verbatim}
(tag) (d/s) (fake#) (file_location) (cross_section) (n_events)
\end{verbatim}
The \texttt{(cross\_section)} and \texttt{(n\_events)} entries are {\em optional}, meaning that they can be excluded from file without causing problems with the operation of Seer. These values are used to scale the per-event weighting. By default, Seer will read these values from the LHCO file directly if they are not set, but a user may wish to rescale the cross section and can do so to override the automatic scan performed by Seer. Further details of the entries in the {\file seer\_files.txt} file are provided in Table \ref{tab:files}.

\begin{table}
\caption{Explanation of the settings for the {\file seer\_files.txt} file.\label{tab:files}}
\noindent\begin{tabularx}{\textwidth}{|>{\bfseries}lX|}
\hline
Parameter & Explanation\\
  \hline
\texttt{(tag)} & This should be an alphanumeric (including symbols ()-+=,.\#\%\/\_). Multiple sequential entries of identical tags will result in these files being grouped together, with cross sections being combined based on the status of the \texttt{(d/s)} flag. Note that the tag choice here will also be the text entry in the legend for that file.\\
\texttt{(d/s)} & This is a flag with two possible values: ``d" or ``s". A choice of ``d" indicates that this file has a different phase space or process from other files with the same tag, and the cross section should be summed. A choice of ``s" indicates that this file duplicates the same phase space as the previous file in the list, and so the total cross section should be averaged between all files with an ``s" and the same tag. \\
\texttt{(fake\#)} & This is a flag for adding in a pseudo-fake rate, and is a number between 0 and 9, inclusive. This is discussed further in Sec. \ref{sec:fakes}.\\
\texttt{(file\_location)} & This is the path (full or relative) to the LHCO file, including the file name.\\
\texttt{(cross\_section)} & This is a cross section value (in pb) that rescales the total cross section for that individual file only. This can be used to rescale total cross sections as desired, or to combine multiple different processes with the same tag, as discussed below.\\
\texttt{(n\_events)} & This is the number of events that corresponds to the cross section value. This entry must accompany the \texttt{(cross\_section)} entry when it is present. Values can be either pre- or post- matching (e.g. the \texttt{(cross\_section)} and \texttt{(n\_events)} settings 5.31 \hspace{1mm} 50000 would be the same as 3.5955 \hspace{1mm} 33856).\\
\hline 
\end{tabularx}\\
\end{table}

To better explain the behaviour of Seer, it is useful to examine some example scenarios to clarify the interplay between the (tag) and (d/s) settings.\\
\vspace{2mm}
\begin{tabularx}{\textwidth}{>{\bfseries}lX}
Scenario 1:& Generated $p p \rightarrow W^+ W^-$ and $p p \rightarrow Z$, listed with a common tag (e.g.  ``background") and with the flag ``d". This results in a single histogram being generated combining the events from both files, and with $\sigma_{tot} = \sigma_{WW}+\sigma_{Z}$.\\
Scenario 2:& Generated 10 files of $p p \rightarrow W^+ W^-$, listed with a common tag and the flag ``s". This results in a single histogram with total cross section averaged over all files ($\sigma_{tot} = \sum \sigma_i \times n_i / \sum n_i$), and the per-event cross section weighted by the inverse of the combined sum of events from all files. Note that these files have to be generated with the exact same phase space limits, otherwise combining them will not be valid.\\
Scenario 3:& Generated 10 files of $p p \rightarrow \ell^+ \ell^-$ each with different, non-overlapping ranges of $m_{\ell\ell}$ (such as 0-100, 100-200, 200-300, 300-400, etc...), listed with a common tag and with the flag ``d". This results in a single histogram with the combined total cross section ($\sigma_{tot} = \sum \sigma_i$). This method results in significantly better sensitivity over the full range of phase space as compared with 10 files with identical phase space.\\
\end{tabularx}
\vspace{2mm}
Files with unique tags will always be treated independently, and be plotted with a their unique tag. Thus, if there is only one file for a given tag, the choice of the \texttt{(d/s)} flag is unimportant. 

There is currently no way to automatically combine multiple files of each of two separate processes with the same tag (i.e. Scenario 1 but with multiple files of each of the processes). This is why the cross section and event number overrides are implemented. In order to use this correctly, the user must manually reduce the cross section to the per-event value following the formula:
\begin{equation}
\sigma_{i}^\prime = \displaystyle\frac{\sigma_{i} \times n_{i}}{\displaystyle\sum_{j=1}^k n_{j}}
\end{equation}
where $\sigma_{i}$ is the cross section listed in the LHCO file. If all files have the same number of events, then this works out to simply $\sigma_{i}/k$. 

Determining the per-event cross section is only slightly more complicated when users are employing matching and merging algorithms in the generation stage. Currently, the MG5+aMC@NLO plus Pythia package lists the total cross section before matching and the total events generated before matching/merging, as well as the events after matching/merging, in the LHCO file. In this case, the relationship $\sigma_{before}/n_{before} = \sigma_{after}/n_{after}$ holds (where {\em before}/{\em after} refers to before/after merging/matching), and so the per-event weight can be determined with $\sigma_{before}/n_{before}$, which is calculated at the MG5+aMC@NLO stage.

\clearpage

\subsection{{\file seer\_cuts.txt} \label{sec:cuts}}

In the {\file seer\_cuts.txt} file, four different types of entries are possible: tagging, trigger, ignores and cuts. These will each be addressed separately below. The general format for the {\file seer\_cuts.txt} file is given by:
\begin{verbatim}
Signal ## Do not alter this line!
##TAGGING
(nlow) (nhigh) (type), pt (ptlow) (pthigh), eta (etalow) (etahigh)
Extra ## Do not alter this line!
##TRIGGER
trigger (type) pt (pt1) (pt2) (pt3) (pt4)
##IGNORES
ignore (type) with (pt/eta) between (kinlow) (kinhigh)
ignore (type) with (pt/eta) over (kinlow)
##CUTS
addcut (type) with (variable) between (kinlow) (kinhigh)
addcut (type) with (variable) over (kinlow)
\end{verbatim}
Changeable settings are enclosed in parentheses. The explanations for each section (tagging, trigger, ignore, and cut) are given below, along with the explanation of each relevant setting entry.

\subsubsection{Tagging}

Tagging determines the particle combinations that Seer uses in its calculations. Tagging criteria are added after the ``Signal" entry in the {\file seer\_cuts.txt} file and before the ``Extra" entry. When passing events to the cut section for analysis, only events that meet precisely the criteria in this list will be included.

\begin{table}
\caption{Tagging settings definitions and explanations.\label{tab:tagging}}
\noindent\begin{tabularx}{\textwidth}{|>{\bfseries}lX|}
\hline
Parameter & Explanation\\
  \hline
\texttt{(nlow)} & This is the minimum number of that type of particle.\\
\texttt{(nhigh)} & This is the maximum number of that type of particle.\\
\texttt{(type)} & This is the type of particle - options include:\\
	& ele -- electron\\
	& muo -- muon\\
	& lep -- lepton, includes muon and electrons.\\
	& tau -- hadronic taus\\
	& ljt -- light jets (does not included $b$-tagged jets or hadronic tau)\\
	& bjt -- $b$-tagged jets\\
	& jet -- all hadronic jets (includes hadronic taus, light jets and bjets)\\
	& gam -- photons\\
	& all -- all particle types combined\\
\texttt{(ptlow)} & the minimum amount of transverse momentum (in GeV) for that specific type of particle\\
\texttt{(pthigh)} & the maximum amount of transverse momentum (in GeV) for that specific type of particle (99999 is often used to represent a value large enough that there is no upper cutoff in $p_T$)\\
\texttt{(etalow)} & the minimum pseudorapidity for that specific type of particle (typically negative value)\\
\texttt{(etahigh)} & the maximum pseudorapidity for that specific type of particle (typically positive value)\\
 \hline 
\end{tabularx}\\
\end{table}

The format for an entry is:
\begin{verbatim}
(nlow) (nhigh) (type), pt (ptlow) (pthigh), eta (etalow) (etahigh)
\end{verbatim}
These entries are not inclusive to particles that are not listed in the tag list. The seer process goes through the list of tagging entries and marks off every particle entry in an event as tagged. Once all tagging entries have been identified, the presence of any untagged particle in the event listing will result in the entire event being identified as ``untagged". There are two ways to make an analysis more inclusive (e.g. inclusive for jets) -- add an extra entry of 0-99 of that particle type with a wide range of kinematics, or add an entry to the ignore list (discussed in Section \ref{sec:ignore}). To make it clear how this works, several examples are included below.

\noindent{\bf Example 1:} This example will only perform an analysis on events that contain between 1 and 2 leptons (either muons or electrons) that have $10 < p_T < 99999$ GeV and within $|\eta| < 2.5$, as well as any number of light jets between 0 and 99 that have $10 < p_T < 40$ GeV, and within $|\eta| < 5$. Note, the presence of a $\tau_h$ or $b$-tagged jet of any $p_T$, or a light jet with $p_T>40$ GeV would cause the entire event to be rejected from the analysis.
\begin{verbatim}
##TAGGING
1 2 lep, pt 10 99999, eta -2.5 2.5
0 99 ljt, pt 10 40, eta -5 5
\end{verbatim}

\noindent{\bf Example 2:} This example will include only events with exactly two high $p_T$ light jets ($200 < p_T < 99999$) and then be inclusive to all light jets with $p_T$ below 100 GeV. Note the double entry of the same type of particle. Seer will tag the items on the list in the order they are listed, and in order from high to low $p_T$.
\begin{verbatim}
##TAGGING
2 2 ljt, pt 200 99999, eta -2.5 2.5
0 99 ljt, pt 10 100, eta -5 5
\end{verbatim}

\subsubsection{Trigger}

Triggers are implemented in Seer as minimum thresholds for $p_T$ required for an event to be considered for analysis. These are effectively treated as step-function cuts at the threshold values. All triggers listed in {\file seer\_cuts.txt} are combined, so any given event will be included in the analysis if it passes the requirements for at least one trigger. To effectively include all events, or rather to turn off the trigger, all of the trigger lines should be commented out (double hash tag, \#\#, preceding the lines).

The format for trigger entries is:
\begin{verbatim}
trigger (type) pt (pt1) (pt2) (pt3) (pt4)
\end{verbatim}
The text for ``trigger" and ``pt" are both necessary and are not settings to be changed. The \texttt{(type)} entry is based on the list of possible triggers, not based on the type of particles. The currently implemented types of triggers are listed in Table \ref{tab:triggers}. There are up to four $p_T$ thresholds that can be set. For trigger types of a single particle type (like jets, which can have , this should be the $p_T$ ordered list of the thresholds. For triggers of a combination of types

\begin{table}
\caption{Implemented trigger types, with explanations. \label{tab:triggers}}
\noindent\begin{tabularx}{\textwidth}{|>{\bfseries}lX|}
\hline
Trigger & Explanation\\
\hline
ele  & electron, four numerical options list the $p_T$ ordered trigger thresholds (a value of 0 turns that option off)\\
muo & muon, four numerical options list the $p_T$ ordered trigger thresholds (a value of 0 turns that option off)\\
gam & photon, four numerical options list the $p_T$ ordered trigger thresholds (a value of 0 turns that option off)\\
tau & hadronic tau, four numerical options list the $p_T$ ordered trigger thresholds (a value of 0 turns that option off)\\
met & missing transverse energy, only the first numerical option should list the missing $E_T$ threshold, the rest should be set to 0\\
ljt & light jet, four numerical options list the $p_T$ ordered trigger thresholds (a value of 0 turns that option off)\\
jet & jet, includes $b$ tagged jets, four numerical options list the $p_T$ ordered trigger thresholds (a value of 0 turns that option off)\\
bjt & $b$ tagged jets, four numerical options list the $p_T$ ordered trigger thresholds (a value of 0 turns that option off)\\
tae & single hadronic tau plus electron, numerical option \texttt{(pt1)} is the $p_T$ threshold of the $\tau_h$, numerical option \texttt{(pt2)} is the $p_T$ threshold of the electron (\texttt{(pt3)} and \texttt{(pt4)} should be set to 0)\\
tam & single hadronic tau plus muon, numerical option \texttt{(pt1)} is the $p_T$ threshold of the $\tau_h$, numerical option \texttt{(pt2)} is the $p_T$ threshold of the muon (\texttt{(pt3)} and \texttt{(pt4)} should be set to 0)\\
jat & single jet (includes $b$ tagged jets) plus missing $E_T$, numerical option \#1 is the $p_T$ threshold of the jet, numerical option \texttt{(pt2)} is the $\met$ threshold (\texttt{(pt3)} and \texttt{(pt4)} should be set to 0)\\
tat & single $\tau_h$ plus missing $E_T$, numerical option \texttt{(pt1)} is the $p_T$ threshold of the $\tau_h$, numerical option \texttt{(pt2)} is the $\met$ threshold (\texttt{(pt3)} and \texttt{(pt4)} should be set to 0)\\
eam & single electron and single muon, numerical option \texttt{(pt1)} is the $p_T$ threshold of the electron, and \texttt{(pt2)} is the $p_T$ threshold of the muon (\texttt{(pt3)} and \texttt{(pt4)} should be set to 0)\\
\hline 
\end{tabularx}
\end{table}

\subsubsection{Ignore\label{sec:ignore}}

Seer has an extra feature called ``ignore", which allows fine tuning adjustment of how Seer treats particle entries in an event. This is particularly useful when using a single, inclusive particle type in which there are different kinematic limits for each. For example, the ATLAS detector can detect electrons with pseudorapidities within $|\eta| < 2.47$, and muons within $|\eta| < 2.5$. Asking Seer to tag general leptons within $|\eta| < 2.5$ would nominally include any electrons with $2.47 < |\eta| < 2.5$. Other uses include specifying whether particle information is discarded, as in \cite{atlas_conf_2013_047}, where it is stated that ``Following this step, all jet candidates with $|\eta|>2.8$ are discarded." This indicates that the event is retained even where jet candidates occur outside $|\eta| = 2.8$, but that the information on those jets are not included in calculations of the kinematics that characterize the event. This kind of treatment is not handled in the Delphes detector simulation.

An ignore setting tells Seer to {\em remove} a particle listing that meets that criteria from the event list, and from all sub-lists within Seer, but does not otherwise change the rest of the event. For example, an electron that is ignored is removed from the electron lists, the lepton lists, and the inclusive ``all" list. Since ignores are run before tagging and trigger, as shown in Figure \ref{fig:flowchart}, particles that are ignored will not participate in those processes.

There are two possible entry types for ignore statements, which are:
\begin{verbatim}
ignore (type) with (pt/eta) over (kinlow)
ignore (type) with (pt/eta) between (kinlow) (kinhigh)
\end{verbatim}
where text entries not enclosed in parentheses are text strings that are required and should not be changed. An ``over" statement indicates that any particles with kinematic ($p_T$ or $\eta$) values larger than the (kinlow) listed will be ignored. A ``between" statement indicates that any particles with kinematic ($p_T$ or $\eta$) values between \texttt{(kinlow)} and \texttt{(kinhigh)} will be ignored. In this case, the \texttt{(eta)} is absolute valued. Thus setting \texttt{eta over 2.0} would be $|\eta|>2.0$. Further descriptions of the limits for these settings is listed in Table \ref{tab:ignore}.

\begin{table}
\caption{Description and explanation of options for ignore feature.\label{tab:ignore}}
\noindent\begin{tabularx}{\textwidth}{|>{\bfseries}lX|}
\hline
Parameter & Explanation\\
\hline
\texttt{(type)} & This can be any particle type from the following list: ele, muo, lep, tau, jet, ljt, bjt, gam.\\
\texttt{(pt/eta)} & This is a text string that is either ``pt" or ``eta" and tells Seer what the kinlow and kinhigh values represent.\\
\texttt{(kinlow)} & This is the lower threshold for the ignore statement.\\
\texttt{(kinhigh)} & This is the upper threshold for the ignore statement.\\
\hline 
\end{tabularx}
\end{table}

{\bf Special Note:} If a user is interested in being inclusive to a particular type of particle (assuming the event signature and trigger is based on other particle states), perhaps jets with $p_T< 30$~GeV, this can be done with a tagging statement of e.g. ``0 99 jets, pt 0 30, eta -2.5 2.5". If the user were to calculate the total $H_T$ of events using this method, these low $p_T$ jets would be included in the $H_T$ calculation. Alternatively, if a user were to use an ignore statement of ``ignore jet with pt between 0 30", then events would be similarly inclusive of these jets however all Seer calculations would be blind to these jets -- the low $p_T$ jets would not be included in total $H_T$ calculations or any other calculations that involve jets. Thus, a user needs to be aware of the effect of using ignore versus tagging statement on inclusiveness of events.

\subsubsection{Cuts}

One of the primary functions of Seer is in the application of cuts. Many types of cut options have been implemented by default. The addition of user-defined cuts is discussed in Section \ref{sec:mod}. This section discusses only the default cuts. To clarify possible confusion between cuts and ignores: cuts {\em remove the entire event} if the criteria is met, where as ignores {\em remove only the specific particle listing within the event} if the criteria is met while leaving the rest of the event untouched.

\begin{table}
\caption{Description and explanation of options for addcuts feature.\label{tab:cuts}}
\noindent\begin{tabularx}{\textwidth}{|>{\bfseries}lX|}
\hline
Parameter & Explanation\\
\hline
\texttt{(type)} & This can be any particle type from the following list: ele, muo, lep, tau, jet, ljt, bjt, gam, met, all.\\
\texttt{(variable)} & For all particle lists (ele through gam in the above list), there is a common set of cut variables. For all and met, there are special cut variables. See Tables \ref{tab:kin1} and \ref{tab:kin2}.\\
\texttt{(kinlow)} & This is the lower threshold for the cut statement.\\
\texttt{(kinhigh)} & This is the upper threshold for the cut statement.\\
\hline 
\end{tabularx}
\end{table}

The format for cut entries is:
\begin{verbatim}
addcut (type) with (variable) over (kinlow)
addcut (type) with (variable) between (kinlow) (kinhigh)
\end{verbatim}
This format is very similar to the format for the Ignores listed previously. However, there are many more options available for the \texttt{(type)} and \texttt{(variable)} listings. Further explanations for the constraints on these settings are given in Table \ref{tab:cuts}. Furthermore, the types of \texttt{(variable)} options are dependent on the particle \texttt{(type)} chosen. Details of the kinematic cuts available for each particle type are listed in Tables \ref{tab:kin1} and \ref{tab:kin2}. 

Cuts that do not necessarily have a single particle type, cuts that depend on multiple types of particles, and cuts that are created by the user, require a \texttt{(type)} setting of ``all". The eta, pt, num, ht, im2 and mas cuts, discussed in Table \ref{tab:kin1} are also implemented for ``all". All addition cuts that are also defined only for ``all" are listed in Table \ref{tab:kin2}. The ``met" setting for \texttt{(type)} only has one possible \texttt{(variable)} option, which is ``pt".

The following are a few examples for cuts that can be implemented:\\
\noindent{\bf Example 1:} Removing all events with light jet $p_T<50$~GeV is accomplished with the line \texttt{addcut ljt with pt between 0 50}.\\
\noindent{\bf Example 2:} Removing all events with any combination of dilepton invariant mass within 20 GeV of the $Z$ mass is accomplished with the line \texttt{addcut lep with im2 between 72 112}.\\
\noindent{\bf Example 3:} Removing all events $\met > 50$~GeV is accomplished with the line \texttt{addcut met with pt over 50}.\\

\begin{table}
\caption{Cut variables common to particle (type) selections: ele, muo, lep, tau, jet, ljt, bjt and gam.\label{tab:kin1}}
\noindent\begin{tabularx}{\textwidth}{|>{\bfseries}lX|}
\hline
Cut Type & Explanation\\
\hline
eta & Pseudorapidity of any particle of that type. If any particle of that type has a pseudorapidity value within the specified range, the entire event is cut.\\
pt & Transverse momentum of any particle of that type.  If any particle of that type has a transverse momentum value within the specified range, the entire event is cut. This is the only cut type currently implemented for the ``met" particle type.\\
pt1 & Transverse momentum of only the leading particle of that type. Cuts events based only on the $p_T$ of the highest $p_T$ particle of that type.\\
pt2 & Transverse momentum of only the subleading particle of that type. Cuts events based only on the $p_T$ of the second highest $p_T$ particle of that type.\\
num & Multiplicity of particles of that type.  Cuts based on the total number of particles of that type in the event (this is highly influenced by the use of tagging versus ignores for inclusivity).\\
ht & Scalar sum of the $p_T$ of all particles of that type. If the total scalar sum of the $p_T$ of all particles of that type is within the specified range, the event is cut.\\
im2 & Invariant mass of any combination of two particles of that type (for leptons, the pairing must be opposite sign). If any pairing of particles of that type has an invariant mass within the range, the event is cut.\\
mas & Total invariant mass of all particles of that type. If the total invariant mass of all particles of that type lies within the range, the event is cut.\\
\hline 
\end{tabularx}
\end{table}

\begin{table}
\caption{Cut variables options available only to the (type) ``all".\label{tab:kin2}}
\noindent\begin{tabularx}{\textwidth}{|>{\bfseries}lX|}
\hline
Cut Type & Explanation\\
\hline
mt2 & Stransverse mass variable. This only works if there are exactly 2 particles in the event (a user may use ignores to neglect low energy jets, for example), and will cut the event if there are more than 2 particles. It calculates the $M_{T2}$ variable using the code developed by \cite{Cheng:2008hk}.\\
maa & Leading diphoton mass. This first checks to see if there are at least two photons in the event, and then cuts based on the invariant mass of the leading two photons. There is some overlap in purposes between this and the im2 option in the previous table. For im2, it searches through all possible combinations of diphotons, but will not provide a cut if there is only one diphoton. This will cut the event if there is only one photon, and only looks at the leading two photons.\\
mll & Same as maa but for leptons.\\
mjj & Same as maa but for jets (all types).\\
mtl & Transverse mass of the only lepton (if only one), or the lowest $p_T$ lepton in the case of multiple leptons.\\
mcol & Collinear invariant mass of exactly two leptons (there can be any number of other particles), including taus. This calculates the invariant mass in the assumption that the missing transverse momentum is comprised from neutrinos that are collinear with the two leptons.\\
osl & Opposite sign (OS) leptons. Behaviour depends on the value of kinlow (kinhigh ignored). If kinlow is 0, all events without an OS lepton pairing are cut. If kinlow is 1, all events with an OS lepton pairing are cut.\\
sfl & Same flavour, opposite sign (SFOS) leptons. Behaviour depends on the value of kinlow (kinhigh ignored). If kinlow is 0, all events without an SFOS lepton pairing are cut. If kinlow is 1, all events with an SFOS lepton pairing are cut.\\
ofl & Opposite flavour, opposite sign (OFOS) leptons. Behaviour depends on the value of kinlow (kinhigh ignored). If kinlow is 0, all events without an OFOS lepton pairing are cut. If kinlow is 1, all events with an OFOS lepton pairing are cut.\\
ssl & Same sign (SS) leptons. Behaviour depends on the value of kinlow (kinhigh ignored). If kinlow is 0, all events without an SS lepton pairing are cut. If kinlow is 1, all events with an SS lepton pairing are cut.\\
\hline 
\end{tabularx}
\end{table}

\clearpage

\subsection{{\file seer\_plots.txt}\label{sec:plots}}

Seer can create either standard one-variable histograms, or can plot two-variable histograms. For one-variable histograms, the $y$-axis contains either cross section values (in fb) or a fraction of events, while the $x$-axis contains the binned kinematic variable value. Note that bins values are not divided by bin width, so Seer does not plot $d\sigma/dx$. Two-variable histograms have binned kinematic variables on both the $x$- and $y$-axes. The way that the bin values are plotted in two-variable histograms depends on whether there is one event type added or multiple event types. For a single event type (i.e. a single signal file), the histogram is plotted with a colour mapping for the cross section, with an associated scale. For multiple event types (i.e. signal files plus different backgrounds), Seer plots a pseudo-Dalitz style plot using the {\small\texttt{box}} option for the {\small\texttt{TH2}} class in ROOT.

For one-variable histograms, Seer distinguishes between two different kinds of events: signal and background. Backgrounds are additive, in that each histogram is the sum of the bin values of all smaller cross section histograms, plus the bin values for the cross section of the specified background. The order of histograms is thus from largest to smallest background, and the bin values for the largest cross section is in fact the sum of all backgrounds, so the user does not need to manually combine backgrounds to estimate the total background cross section. Backgrounds are plotted with a solid fill colour.

Files considered to be part of the signal are not additive, but are re-ordered to plot from largest total cross section to smallest. In addition, signal histograms do not use a fill colour and are only a contour, and are plotted on top of all the backgrounds. Signals are distinguished from backgrounds based on the text string in the ``tag" entry in the {\file seer\_files.txt} file.

The relevant entries in the {\file seer\_plots.txt} file are, where adjustable settings are enclosed in parentheses:
\begin{verbatim}
Disable = (disable)
PlotType = (plotnum)
PlotType2 = (plotnum2)
PlotMaxX = (xmax)
PlotMinX = (xmin)
PlotMaxY = (ymax)
PlotMinY = (ymin)
NumBinX = (nbins)
NumBinY = (nbins)
ChooseLn = (log)
Normalize = (norm)
PlotTxt = (name_string)
SigPref = (signal_string)
FileType = (file_type)
\end{verbatim}
The explanations for each of the entries in this file are given in Table \ref{tab:plots}.

\begin{table}
\caption{Plotting settings definitions and explanations.\label{tab:plots}}
\noindent\begin{tabularx}{\textwidth}{|>{\bfseries}lX|}
\hline
Parameter & Explanation\\
\hline
\texttt{(disable)} & This is a flag to disable the plotting features of Seer (1 = disable, 0 = enable). The text output is still produced. Seer has a maximum of 50 histograms (distinct \texttt{(tag)} entries in {\file seer\_files.txt}), and this flag will necessarily need to be implemented if this number is exceeded.\\
\texttt{(plotnum)} & This is the kinematic variable that is plotted on the $x$-axis. The list of possible kinematic variables is included as comments in the {\file seer\_plots.txt} file. The list is too long to include in this document.\\
\texttt{(plotnum2)} & This is the kinematic variable that is plotted on the $y$-axis, similar to \texttt{(plotnum)}. A value of 0 turns this feature off, and Seer plots a one-variable histogram.\\
\texttt{(xmax)} & This is the maximum value on the $x$-axis.\\
\texttt{(xmin)} & This is the minimum value on the $x$-axis.\\
\texttt{(ymax)} & This is the maximum value on the $y$-axis, which is cross section in units of fb. (Future functionality will include typing ``auto" for this setting.)\\
\texttt{(ymin)} & This is the minimum value on the $y$-axis, which is cross section in units of fb. (Future functionality will include typing ``auto" for this setting.)\\
\texttt{(nbinx)} & This is the number of bins to plot between \texttt{(xmin)} and \texttt{(xmax)}.\\
\texttt{(nbiny)} & This is the number of bins to plot between \texttt{(ymin)} and \texttt{(ymax)}.\\
\texttt{(log)} & Only considers values of 0 or 1. A value of 1 sets the $y$-axis or $z$-axis to use a logscale. A value of 0 sets the $y$-axis or $z$-axis to use a linear scale.\\
\texttt{(norm)} & Only considers values of 0 or 1. A value of 1 normalizes the binned events to the total binned cross section, and the $y$-axis ($z$-axis) contains event fraction values. A value of 0 sets the $y$-axis ($z$-axis) to plot cross section values.\\
\texttt{(name\_string)} & This adds a text string to the output file names. Useful for distinguishing between subsequent runs with identical files, but different settings.\\
\texttt{(signal\_string)} & This is the identifying text of the prefix of the signal file tags. For example, if the tag line in {\file seer\_files.txt} for all signal files is ``sig", this setting should be set to ``sig". All such files will be plotted differently, as discussed above.\\
\texttt{(file\_type)} & This tells Seer the file extension for the produced histogram. Common options are ``pdf", ``eps" and ``png". Possible options are the same as the types of files known to ROOT.\\
\hline 
\end{tabularx}
\end{table}

\clearpage

\subsection{{\file seer\_fakes.txt}\label{sec:fakes}}

For event simulators, fake rates are typically handled internally. In Delphes, there is a probability that a jet will fake an electron, for example. Fake backgrounds can have event rates similar to that of rare signals if there is a combination of a very large background with a very small fake rate. Estimating the cross section that overlaps with the signal can be very challenging using events from the detector simulator; if the fake rate is 1/1000, then out of 50000 events generated, approximately 50 events will be useful. Generating enough statistics thus takes a long time.

To improve on this, Seer has a fake-rate multiplier implemented. In effect, it allows for the possibility of using a much larger event fraction while simulating the effect of the fake rate as a re-scaling of the overall cross section. For example, if a user is interested in the contribution of $W+$jets to $e^+e^-$ signals, and has generated events of $W+nj$, the cross section can be rescaled by the fake rate, and one of the jets can be randomly selected to be added to the electron lists.

{\bf Note: This can over-estimate the fake background if the fake rate is included in both the detector simulator and Seer. Thus, if Seer is going to handle fake rates, it is required that the fake rate be turned off in the detector simulator.}

Control over this process is handled by the {\file seer\_fakes.txt} file. This file is structured as follows, where all setting entries are enclosed in parentheses:
\begin{verbatim}
Rates
(rstate) (fstate) (efficiency)
Schema
scheme 0
scheme 1
addreal (nlow) (nhigh) (rstate), pt (ptlow) (pthigh), eta (etalow) (etahigh)
addfake (rstate) (fstate), pt (ptlow) (pthigh), eta (etalow) (etahigh)
scheme 2
scheme 3
scheme 4
scheme 5
scheme 6
scheme 7
scheme 8
scheme 9
\end{verbatim}
There are 10 total fake schemes that can be implemented. It is recommended that scheme 0 be left empty, as a ``no fakes" option. Note that the scheme number is the same number that should accompany the file in {\file seer\_files.txt}. These setting entries are defined in Table \ref{tab:fake}. There can only be one fake rate defined (under Rates) for each combination of \texttt{(rstate)} and \texttt{(fstate)}, but there can be multiple entries of the same combination of \texttt{(rstate)} and \texttt{(fstate)} as an addfake.

\begin{table}
\caption{Fake rate setting definitions and explanations.\label{tab:fake}}
\noindent\begin{tabularx}{\textwidth}{|>{\bfseries}lX|}
\hline
Parameter & Explanation\\
\hline
\texttt{(rstate)} & This is a real particle state from the list of lep, ele, muo, bjt, gam, tau, ljt.\\
\texttt{(fstate)} & This is a fake particle state, where the possibilities are limited depending on the corresponding rstate. Real light jets (ljt) can fake ele, tau, bjt and gam. Real photons (gam) can fake ele. Real electrons (ele) can fake photons.\\
\texttt{(efficiency)} & This is the probability of the rstate faking the fstate.\\
\texttt{(nlow)} & This is the minimum number of a real particle\\
\texttt{(nhigh)} & This is the maximum number of a real particle\\
\texttt{(ptlow)} & This is the minimum $p_T$ of a real particle\\
\texttt{(pthigh)} & This is the maximum $p_T$ of a real particle\\
\texttt{(etalow)} & This is the minimum $\eta$ of a real particle (etalow = -etamax for symmetric)\\
\texttt{(etahigh)} & This is the maximum $\eta$ of a real particle\\
\hline 
\end{tabularx}
\end{table}

The ``addreal" entries list the requirements needed for Seer to consider the faking of an event. In the example of $W+nj$ faking $e^+e^-$, the assumption is that one electron comes from the decay of the $W$ and the other one is a faked light jet. Thus, the user would set \texttt{addreal 1 1 ele, pt 10 99999, eta -2.5 2.5}, which will look for any event that already has exactly one electron (if this is the only addreal, then the other states are inclusive) within the specified kinematic range. If one is found, then Seer will pass the event along to determine the fake particle.

The ``addfake" entries determines how Seer fakes the state. First, Seer will look among all of the \texttt{(rstate)} particles in the event and identify which ones meet the kinematics listed. It will then randomly select a particle from that list to turn into the \texttt{(fstate)} type, removing the particle from the appropriate lists and adding it to the fake state list(s). In the example of $W+nj$ faking $e^+e^-$, the user would set \texttt{addfake ljt ele, pt 10 99999, eta -2.5 2.5}, which will take one of the light jets from within that kinematic range and turn it into an electron. In the case of an electron, the charge of the electron is determined randomly, as well. Thus, it is possible that a fake electron from $W+nj$ will result in either same sign electrons or opposite sign electrons.

It is important to note the event weight for {\bf all} events within the file, whether a faked particle is present in the event or not, is scaled by the product of all the fake rate efficiencies for each ``addfake" entry in the scheme being used for that file. Continuing with the same example, if the fake rate of jets as electrons is 0.0001, then any file that uses that scheme will rescale the total cross section by 0.0001. Thus, any instances of $e^+e^-$ within those files will have their weight rescaled. This is why processes that fake the signal need to be separated from processes that produce the signal when using the Seer faking system, and why fake rates need to be turned off at the detector simulator level.

\subsection{{\file seer\_settings.txt}\label{sec:settings}}

There are a number of constants that Seer uses for various calculations. The numerical value of these settings can be adjusted in the {\file seer\_settings.txt} file. The format for this file is:
\begin{verbatim}
mlsp = (mass_lsp)
wmas = (W_mass)
wwid = (W_width)
zmas = (Z_mass)
zwid = (Z_width)
tmas = (t_mass)
twid = (t_width)
jeer = (jE_smear)
jmer = (jM_smear)
jher = (jh_smear)
metA = (MET_scale)
metP = (MET_pileup)
metM = (MET_minbias)
\end{verbatim}
Further information about these settings, as well as the default values for them, is listed in Table \ref{tab:settings}.

\begin{table}
\caption{Explanations and default values for the \label{tab:settings}}
\noindent\begin{tabularx}{\textwidth}{|>{\bfseries}llX|}
\hline
Parameter & Default & Explanation\\
  \hline
\texttt{(mass\_lsp)} & 0 & sets the value of the mlsp variable (mass of the LSP setting used for MT2 calculations)\\
\texttt{(W\_mass)} & 80.385 & sets the value of the wmass variable (currently not used)\\
\texttt{(W\_width)} & 2.085 & sets the value of the wwidth variable (currently not used)\\
\texttt{(Z\_mass)} & 91.1876 & sets the value of the zmass variable (currently not used)\\
\texttt{(Z\_width)} & 2.4952 & sets the value of the zwidth variable (currently not used)\\
\texttt{(t\_mass)} & 173.21 & sets the value of the tmass variable (currently not used)\\
\texttt{(t\_width)} & 2.00 & sets the value of the twidth variable (currently not used)\\
\texttt{(jE\_smear)} & 0.00 & sets the value of the jet\_E\_error variable (used to add additional $p_T$ smearing)\\
\texttt{(jM\_smear)} & 0.00 & sets the value of the jet\_m\_error variable (used to add additional jet mass smearing)\\
\texttt{(jh\_smear)} & 0.00 & sets the value of the jet\_eta\_error variable (used to add additional $\eta/\phi$ smearing)\\
\texttt{(MET\_scale)} & 0.75 & sets the value of the met\_scale variable (used to scale MET smearing)\\
\texttt{(MET\_pileup)} & 30 & sets the value of the met\_pileup variable (average \# of pileup events for MET smearing)\\
\texttt{(MET\_minbias)} & 20 & sets the value of the met\_min\_bias variable in GeV (minimum bias energy for MET smearing)\\
\hline 
\end{tabularx}
\end{table}

The $M_{T2}$ calculator included with Seer requires an initial guess of the mass of the LSP particle, which is set in this file. The $W$, $Z$, and $t$ mass settings are currently not used within any calculations. However, their inputs may be useful for any tagging or kinematics that depend on this values. An example where this may be useful would be the cuts employed in the ATLAS tri-lepton search for supersymmetry, where the phase space is separated into bins based on the dilepton invariant mass being smaller, larger or within a window about the $Z$ pole mass.

Additionally, Seer allows for the possibility of adding greater smearing to the jets and missing transverse energy values. All smearing values are calculated using a Gaussian distributed random value centred at zero with widths dependent on the type of jet being smeared. For hadronic jets (applies to light jets, $b$ jets and $\tau_h$ jets), additional jet $p_T$ will be smeared with a width of $\sigma_E=(jE\_smear)\times p_T^j$, jet mass will be smeared with a width $\sigma_m = (jM\_smear)\times m_j$ (useful particularly for large $R$ fat jets), and the $\eta$ and $\phi$ values will be independently smeared with a width $\sigma_R = (jh\_smear)\times \eta(\phi)$ (where \texttt{(jE\_smear)}, \texttt{(jM\_smear)} and \texttt{(jh\_smear)} are the relevant quantities from {\file seer\_settings.txt}). For missing transverse energy, smearing is performed independently on the $x$ and $y$ components with a width given by the formula from \cite{ATL-PHYS-PUB-2013-004}:
\begin{equation}
\sigma_{\not\!E_T} = A(0.40+0.09\sqrt{P})\sqrt{\sum E_T +P\times M}
\end{equation}
where $A$ is an overall scaling factor given in Seer by the variable \texttt{metA} (used to account for existing $\met$ smearing in the detector analysis), $P$ is the average number of pileup events given in Seer by the variable \texttt{metP}, and $M$ is a minimum bias energy given in Seer by the variable \texttt{metM}. The sum of visible transverse energies should be in GeV, as should the value of $M$. The numeric factors are fitted values from \cite{ATL-PHYS-PUB-2013-004}.

\subsection{Command Line Input/Output}

Seer is run using the command {\file ./seer.exe} in the main folder. The seer terminal output provides most of the same information as is in the text file output with a few minor differences. The differences are:
\begin{itemize}
\item The descriptions of the settings files are more descriptive, rather than an identical copy of the lines in the file. 
\item Individual file results are not listed, only group summaries and the overall summary is listed.
\item The histogram information is not listed.
\end{itemize}
The information in the text output file is meant to be useful for recreating the analysis, while the terminal output is meant to be a more descriptive explanation and summary of the analysis.

There are three possible arguments that can be input (individually or in any combination) which will force Seer to run with extra text output to the terminal (not to the output text file). These three arguments represent three modes:
\begin{itemize}
\item debugmode - used to print out information about nearly every step of the analysis process. This is useful to use when Seer crashed mid-run to determine where the problem lies.
\item rejectmode - used to print out a detailed output of all particles within an event {\em only} when that event is rejected. This is useful for analyzing each event that is cut to determine if the cuts are working as intended.
\item verbosemode - turns on output of the individual file summaries into the terminal output, and output of an initial verification of the cross section and event counts in the LHCO files.
\end{itemize}
Note that the debugmode and rejectmode modes produce an excessive amount of text, and it is recommended that they be used only in conjunction with a test LHCO file that contains around 10 events. Creating such an LHCO file is easy and does not require modifying the cross section or event count entries within the file. Such a test file can be created by making a copy of any of the signal or background files being examined and highlighting and deleting all but the first 10 events. (Note: deleting the extra carriage return at the end of the LHCO file can result in problems within Seer.)

\section{Modifying Seer\label{sec:mod}}

There are two aspects of Seer that a user may wish to modify. The first is user-defined cuts, and the second is user-defined plot variables. Both of these require an understanding of how the information is stored in Seer, and then an understanding of how that information will be interpreted and used within Seer. This will be discussed in this section.

\subsection{Accessing information in {\file calculations.cpp}}

Seer stores the information for each particle within a series of vectors of the TLorentzVector class in ROOT. This information is stored only for the current event, and is cleared upon reading in a new event. The following vectors of particles are available:\\
\begin{lstlisting}
std::vector<TLorentzVector> v_gam; // photons
std::vector<TLorentzVector> v_jet; // all jets (includes b-tagged and hadronic tau jets)
std::vector<TLorentzVector> v_bjt; // only b-tagged jets
std::vector<TLorentzVector> v_ljt; // only non-b-tagged and non-hadronic tau jets
std::vector<TLorentzVector> v_tau; // only hadronic taus
std::vector<TLorentzVector> v_lep; // both electrons and muons
std::vector<TLorentzVector> v_lne; // only negatively charged leptons (both electrons and muons)
std::vector<TLorentzVector> v_lpo; // only positively charged leptons (both electrons and muons)
std::vector<TLorentzVector> v_ele; // all electrons
std::vector<TLorentzVector> v_ene; // only negatively charged electrons
std::vector<TLorentzVector> v_epo; // only positively charged electrons
std::vector<int> v_ele_charge; // the charge of electrons in v_ele, in the same order as v_ele
std::vector<TLorentzVector> v_muo; // all muons
std::vector<TLorentzVector> v_mne; // only negatively charged muons
std::vector<TLorentzVector> v_mpo; // only positively charged muons
std::vector<int> v_muo_charge; // the charge of muons in v_muo, in the same order as v_muo
std::vector<TLorentzVector> v_met; // missing transverse energy
std::vector<TLorentzVector> v_all; // all particles, except missing tranverse energy
std::vector<int> v_lep_charge;	// the charge of leptons in v_lep, in the same order as v_lep
\end{lstlisting}
There are six types of particle entries in LHCO files, based on the number in the second column for the particle listing (referred to as tag). These options are 0 (photon), 1 (electron), 2 (muon), 3 (hadronic tau), 4 (jet), and 6 (missing transverse energy). Furthermore, the sign of the entry for number of tracks indicates the charge for electrons and muons, and the b-tag field indicates whether a jet is a b-tagged jet. This is the information that is used to determine which vectors to add the particle entries. These vectors are {\em not} exclusive, such that almost every particle entry exists in multiple lists. The reason for this methodology is to make modifying and adding new calculations more easy for users, at the expense of being more memory intensive.

Users have full access to the TLorentzVector functions for each of the vectors of these objects. The list of functions is available on \url{https://root.cern.ch/root/html/TLorentzVector.html}.

Note: If the user defines their own vector class object, this should be cleared within the {\small\texttt{event\_destroy()}} function in {\file calculations.cpp}.

\subsection{User defined kinematic variables}

The following steps will walk users through the process of adding code to calculate a new kinematic variable to Seer.

\begin{enumerate}
\item[Step 1:] The function needs to be defined within {\file calculations.h}. For an example, a calculation of the leading dijet plus leading lepton invariant mass is discussed through the rest of these steps. The function could be defined with the code:
\begin{lstlisting}
virtual double calc_Mjjl();
\end{lstlisting}

\item[Step 2:] Following this, the function needs to be defined within {\file calculations.cpp} after the text ``This area is for user defined functions." Templates are provided in the code to understand the format. Continuing with the example, the code could be:
\begin{lstlisting}
double event::calc_Mjjl() {
	if ( v_jet.size() >= 2 && v_lep.size() >=1 ) {
		return (v_jet.at(0)+v_jet.at(1)+v_lep.at(0)).M();
	}
	return -99999;
}
\end{lstlisting}
In this example, a few important things should be noted:
\begin{enumerate}
\item The function must first check that the jet and lepton vectors have the appropriate size or the user will have errors during calculation that will stop Seer from functioning. 
\item The particle vectors are ordered from largest to smallest $p_T$, and {\small\texttt{v\_jet.at(0)}} and {\small\texttt{v\_jet.at(1)}} are the leading and subleading jet. For other possible combinations of jets from a larger list of jets, the user will need to define boolean conditions to select the desired jets from the list.
\item This example does not distinguish between b-tagged jets, hadronic tau jets, and light jets. To use only light jets, the user could instead use {\small\texttt{v\_ljt}} instead of {\small\texttt{v\_jet}}.
\item Returning a value of 0 if there are insufficient jets and leptons to calculate a mass can result in the 0th bin being incorrectly filled with events that should not be plotted. Using a value that is unreasonable for the plotting scale will ensure that such events do not appear on the histogram.
\end{enumerate}

\item[Step 3:] To add plotting capabilities, allowing this new function as a plotting variable, two changes must be made. The first is to add the function to the list of variables output by the function {\small\texttt{calc\_ps(int hist\_type)}} within {\file calculations.cpp}. The function needs to be added to the end of the list of else-if statements with an incremented reference number (see Section \ref{sec:plots} for more details on the reference number, \texttt{(plotnum)}). Continuing with the example, the code could look like:
\begin{lstlisting}
else if ( hist_type == 49 ) {
	event_ps = calc_Mjjl();
}
\end{lstlisting}

\item[Step 4:] The second change to allow plotting requires modifying the {\file plots.cpp} file. The area to modify, along with a template, is provided following the text string ``USER AREA." Using the same \texttt{(plotnum)} value from {\small\texttt{calc\_ps()}}, the {\small\texttt{name\_text}}, {\small\texttt{title\_text}}, {\small\texttt{yaxis\_text}} and {\small\texttt{xaxis\_text}} strings must be modified as needed.  The {\small\texttt{name\_text}} variable is a string that will be added to the filename of the output files (see Section \ref{sec:plots} for more information). The {\small\texttt{title\_text}} variable is a string that will nominally be added as a title to the ROOT canvas, however this is currently disabled in the style file, {\file style.h}. The {\small\texttt{yaxis\_text}} and {\small\texttt{xaxis\_text}} are ROOT-LaTeX strings that will become the labels on the y and x axes, respectively. Currently, the $y$-axis in Seer output plots is only cross section values in units of fb, so the {\small\texttt{yaxis\_text}} variable does not need to be changed. Continuing with the example, this code could look like:
\begin{lstlisting}
else if ( plot_type == 49 ) {
	name_text = "mjjl";
	title_text = "Dijet+Lepton Mass";
	yaxis_text = "#sigma (fb)";
	xaxis_text = "M_{jjl} (GeV)";
}
\end{lstlisting}
\end{enumerate}

At this point, all necessary changes have been made to create plots of with this variable. The number 49 is now an accessible \texttt{(plotnum)} in {\file seer\_plots.txt}. This kinematic variable can also be implemented for use as a cut calculation, but this is a more nuanced issue that will be discussed in the next section.

\subsection{User defined cuts}

User defined cuts can be implemented using two methods. The first is using a boolean function that cuts based on some (complex) criteria. The second is by using a kinematic function. The key to implementing cuts is in setting {\bf both} the {\small\texttt{cuttest}} and {\small\texttt{intcuttest}} variables to be true if the condition is met. The logic used in Seer is that if {\small\texttt{cuttest}} is TRUE, the event will {\bf not} be included in the analysis, and it will not be passed to the plotting function.

The variable {\small\texttt{intcuttest}} is reset for each run through the different cuts implemented, distinguishing which cut settings eliminate the event (since an event may be cut by multiple cut conditions). Alternatively, {\small\texttt{cuttest}} is not reset through each cut, and so it simply represents an overall test whether the event is cut. This is why both variables must be set to be TRUE.

The following steps outline the procedure to define a boolean function to Seer.

\begin{enumerate}
\item[Step 1:] The function needs to be defined within {\file calculations.h}. The naming scheme for boolean functions used for cuts is a prefix of {\small\texttt{cut\_}}. As an example, a function that cuts based on number of jets with a complex combination of $p_T$ thresholds is discussed through the rest of these steps. The function could be defined with the code:
\begin{lstlisting}
virtual bool cut_jets();
\end{lstlisting}
Note that this function does not make use of the \texttt{(kinlow)} and \texttt{(kinhigh)} variables from {\file seer\_cuts.txt}, and is therefore a hard coded function. Users may wish to access information from \texttt{(kinlow)} and \texttt{(kinhigh)} for a more nuanced cut. To do this, the user could define the function as:
\begin{lstlisting}
virtual bool cut_jets(double kin_low, double kin_high);
\end{lstlisting}
If the user expects integer input for different cases, based only on the value of \texttt{(kinlow)}, the function could be defined as:
\begin{lstlisting}
virtual bool cut_jets(int kin_low);
\end{lstlisting}
Note that this would require recasting the double valued {\small\texttt{kin\_low}} variable as an integer at the point of use of the function. The following steps assume that the second of these three function definitions is implemented.

\item[Step 2:] Following this, the function needs to be defined within {\file calculations.cpp} after the text ``This area is for user defined functions." Templates are provided in the code to understand the format. Continuing with the example, the code could be:
\begin{lstlisting}
bool event::cut_jets(double kin_low, double kin_high) {
	if ( v_jets.size() >= 4  && v_tau.size() == 0 ) {
		if ( v_jet.at(0).Pt() > kin_high && v_jet.at(1).Pt() > kin_low && v_jet.at(3).Pt() > 50 ) {
			return true;
		}
	}
	return false;
}
\end{lstlisting}
For this code, a few important things should be noted:
\begin{enumerate}
\item This function requires more than 4 jets with 0 tau jets, thus it is inclusive to b-tagged jets and light jets, but not hadronic taus.
\item This function requires that the leading jet has a $p_T$ larger than the value of kin\_high, and a subleading jet with a $p_T$ larger than the value of {\small\texttt{kin\_low}}.
\item This function requires that the fourth jet ({\small\texttt{v\_jet.at(3)}}) has a $p_T>50$~GeV. Since the jet list is $p_T$ ordered, this also means that the third jet ({\small\texttt{v\_jet.at(2)}}) will have a $p_T>50$~GeV.
\end{enumerate}

\item[Step 3:] Following the definition of the function, the cut must be implemented within the {\small\texttt{run\_cut()}} function in {\file calculations.cpp}. Templates have been provided following the text string ``This area is for user defined cuts." Adding the cut first requires defining a unique string that will be used to identify the cuts in {\file seer\_cuts.txt}, which will be tested against the variable {\small\texttt{kin\_type}}. The second issue requires understanding exactly the condition of the function which will result in cuts. This will be more clear in the context of continuing the example, which can be implemented with the code:
\begin{lstlisting}
if ( kin_type == "myjet"  ) {
	if ( !cut_jets(kin_low,kin_high) ) {
		if ( debugmode ) {
			std::cout << "Event rejected because of a myjet cut" << std::endl;
		}
		cuttest = true;
		intcuttest = true;
	}
}
\end{lstlisting}
A few important things should be noted:
\begin{enumerate}
\item This script sets {\small\texttt{cuttest}} to TRUE if the return of the {\small\texttt{cut\_jets()}} function is FALSE. This is an important distinction. Thus, events that don't have jets falling within the kinematic constraints defined in the function will be removed. Close attention to the logic is necessary to properly code these cuts. When {\small\texttt{cuttest}} is TRUE, the event is removed.
\item The {\small\texttt{debugmode}} feature has been implemented to address the complexity of these cuts. Running in {\small\texttt{debugmode}} on a reduced sample of events ($\sim10$ events is a good number) is useful for testing the behaviour.
\item This is not optimized coding, written in order to make understanding the behaviour easier.
\end{enumerate}
\end{enumerate}

The procedure for performing a cut on user defined variables is very similar. Steps 1 and 2 from the previous section need to be followed to create the kinematic calculation function. Following this, Step 3 from this section outlines the basic procedure to implement the cut. There is a template provided within the function {\small\texttt{run\_cut()}} in {\file calculations.cpp}. For the example of the dijet plus lepton invariant mass function defined in the previous section, the following code would implement a cut based on the {\small\texttt{kin\_low}} and {\small\texttt{kin\_high}} values:
\begin{lstlisting}
if ( kin_type == "JJL"  ) {
	if ( v_jets.size() >= 2 && v_lep.size() >= 1 ) {
		testvalue = calc_Mjjl();
		if ( testvalue > kin_low && testvalue < kin_high ) {
			if ( debugmode ) {
				std::cout << "Event rejected because of a JJL cut of any particles with " << testvalue << std::endl;
			}
			cuttest = true;
			intcuttest = true;
		}
	}
	else { // This else statement will result in removal of all events that do not have at least 2 jets and at least 1 lepton.
		cuttest = true;
		intcuttest = true;
	}
}
\end{lstlisting}
In this code, the calculated kinematic variable is passed to the {\small\texttt{testvalue}} variable. This is to save on computation time, which is especially important for more complex codes like the stransverse mass, {\small\texttt{calc\_MT2()}, and razor, {\small\texttt{calc\_MR()}, functions, which would add significant computing time to compute once for each of the comparisons to kin\_low and kin\_high. Also note that the dijet plus lepton mass can only be calculated on events with at least two jets and at least one lepton, so an extra logical test has been added to cut events that don't meet these minimum requirements.

In either of the two cases presented, adding these functions in the region outlined opens up the ability to add addcut lines to the {\file seer\_cuts.txt} file such as:
\begin{verbatim}
addcut all with myjet between 300 400
addcut all with JJL between 0 500
\end{verbatim}
Based on the definition of the myjet cut, the first line would remove event if the there were fewer than 4 jets, the leading jet had a $p_T < 400$~GeV, the subleading jet had a $p_T<300$~GeV, or if the third and fourth leading jets had $p_T < 50$~GeV. Any or multiple of those conditions would result in removal of the event from the analysis. The JJL cut is more straightforward, as it will remove events in which the dijet plus lepton invariant mass is between 0 and 500~GeV.

\subsection{Other modifications to Seer}

Users may wish to make other modifications to Seer. A few of these changes will be addressed here.

\subsubsection{Changes to histogram colour schemes}

Currently, Seer checks the number of histograms to plot and chooses from a few spectral choices. For a different colour scheme, users can alter the settings within the constructor of the {\small\texttt{plot}} class within {\file plots.cpp}. The colours available in ROOT are listed at \url{https://root.cern.ch/root/html/TColorWheel.html}. The three different colour scenarios are for 1-9 histograms, 9-15 histograms, and more than 15 histograms. It is not recommended that users plot more than 15 histograms, however, except in very specialized situations, as the lines plots will become increasingly busy and difficult to read. However, Seer can effectively handle up to 50 overlapping histograms.

\subsubsection{Changing canvas/axis settings}

The canvas and axis for the output histogram is set at the end of the constructor of the {\small\texttt{plot}} class within {\file plots.cpp}. This area is labeled by a comment string stating ``Canvas/Axis settings". This area controls the offset and size of the axis labels, as well as the axis tick settings. Some default settings also exist within the {\file style.h} file. Modifications to {\file style.h} should be handled by users who are very experienced with ROOT. See the ROOT documentation on \url{root.cern.ch} for more information on each of these settings.

\subsubsection{Changing the legend}

The upper right corner of the legend defaults to the upper-right-most corner of the canvas, and has a width of 19\% of the canvas. The height of the legend is the larger value of 20\% or $3.5\% \times n_{hist}$, where $n_{hist}$ is the number of histograms included. Changes to the legend location and size can be made at the start of the {\small\texttt{generate\_plot()}} function within the {\small\texttt{plot}} class in {\file plots.cpp}.

\section{Examples}

\subsection{Analysis Prototyping and Data Visualization}

One of Seer's primary functions is for analysis prototyping and data visualization, where kinematic distributions of signals and backgrounds can be explored with minimal effort. By adjusting a few settings files, many different tagging permutations and cuts can be implemented and visualized, producing publication quality figures.

In a recent study of light pseudoscalar production at LHC13 \cite{Kozaczuk:2015bea}, Seer was used to analyze SM and beyond the SM production of dilepton events in association with $b$ jets generated with MG5+aMC@NLO. A total of 40.7 million events were generated, and heavy states ($W$, $Z$, $a$, $t$) were decayed democratically over all lepton generations ($e$, $\mu$, $\tau$). Intermediate figures from this study are included in Figure \ref{fig:mumu}, which shows di-muon event production at the LHC with at least 1 $b$-tagged jet, and inclusive to light jets with $p_T<40$~GeV. Backgrounds of $\gamma^*/Z + 0-2 b + 0-2j$, $W^+W^- + 0-2b + 0-2j$, $W+0-2b+0-2j$ and $WZ+0-2b+0-2j$ are included, as well as for a BSM pseudoscalar with an enhanced coupling to down-type fermions over SM Higgs couplings, for masses of the pseudoscalar $m_a = 50,80$~GeV. Dimuon decays arise from direct dimuon production, but also as decay products from $\tau^+\tau^-$ production.

\begin{figure*}[tb]
\centering
\mbox{\includegraphics[width=0.45\textwidth]{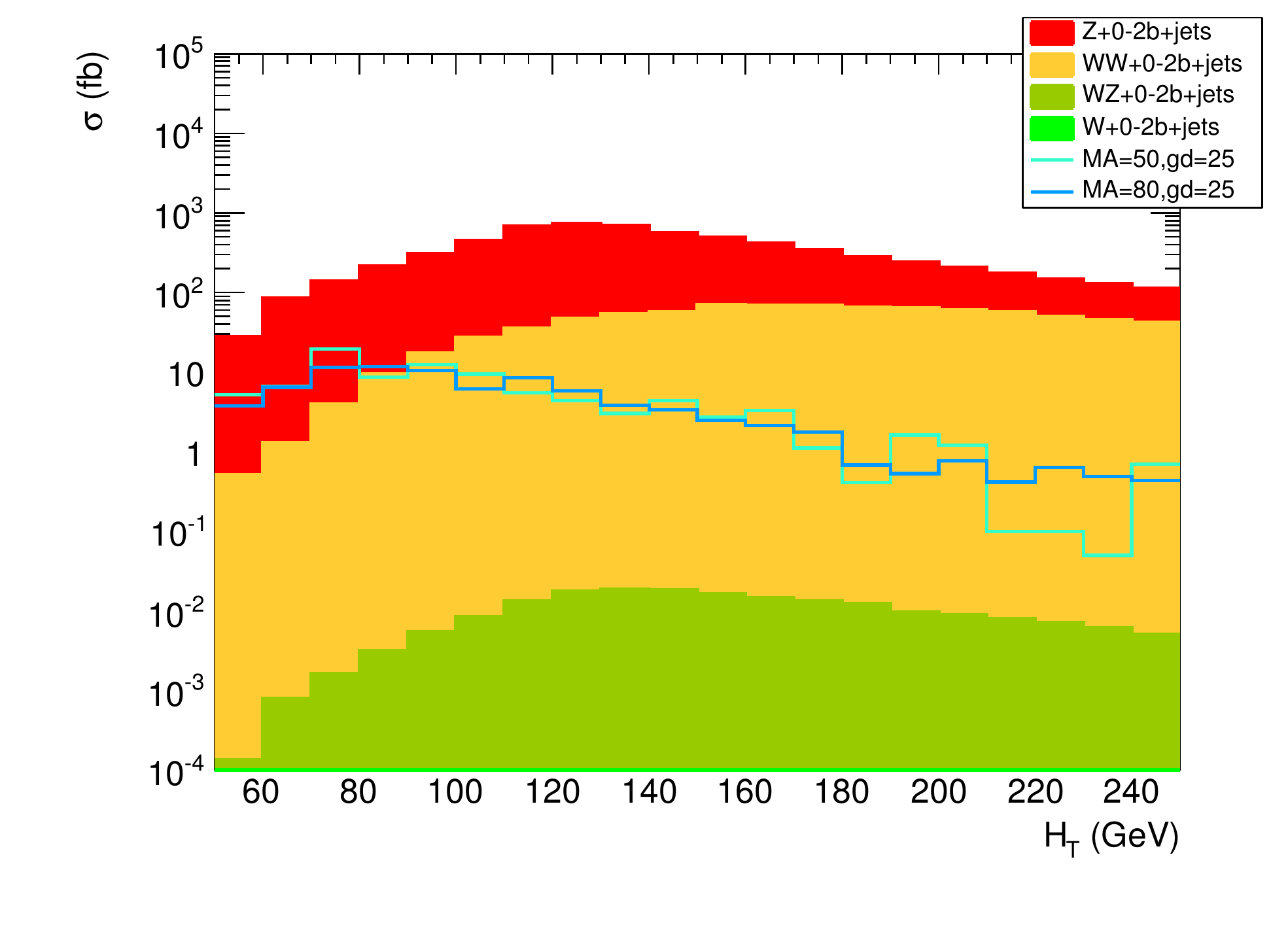}
\includegraphics[width=0.45\textwidth]{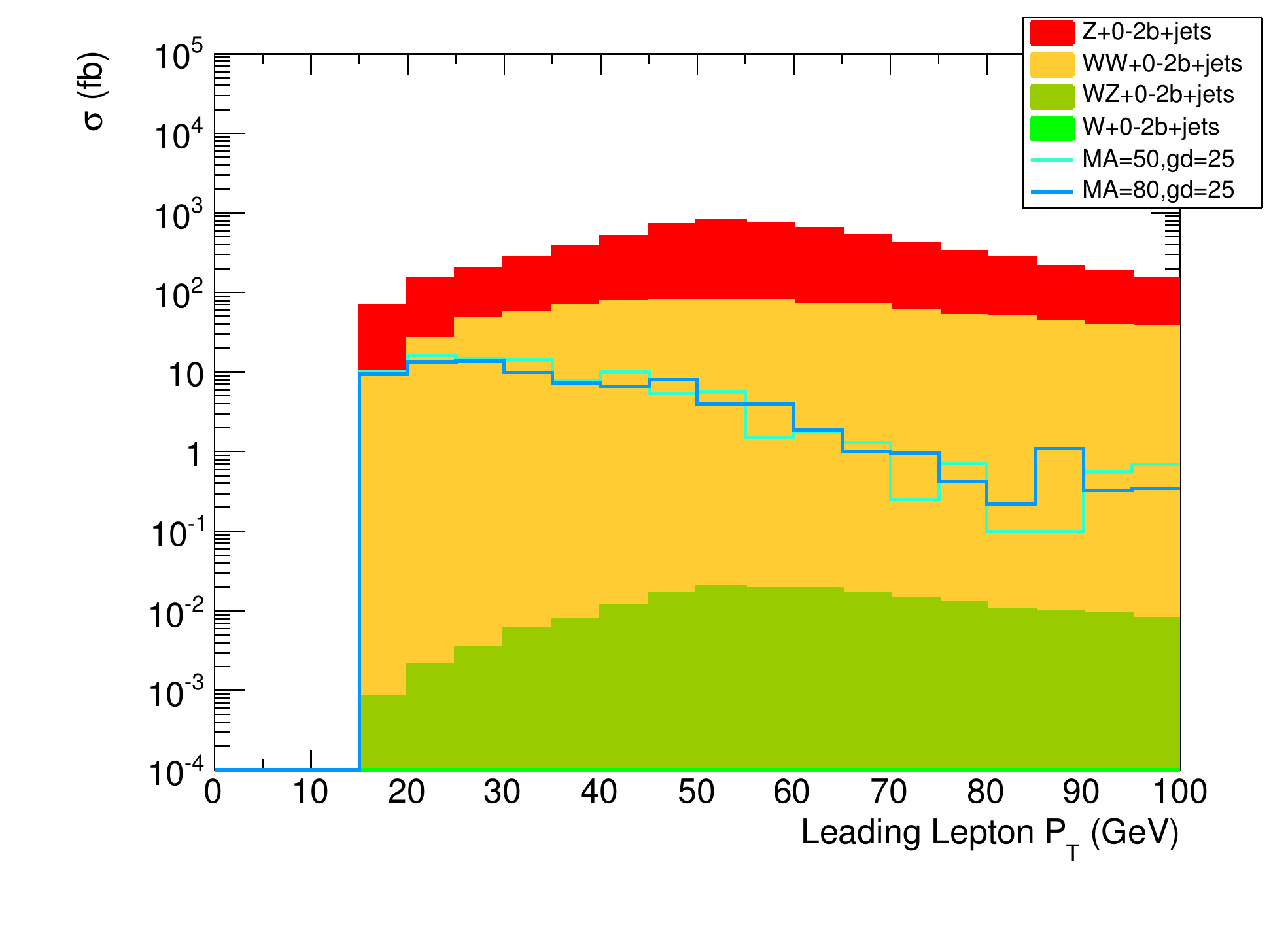}}\\
\mbox{
\includegraphics[width=0.45\textwidth]{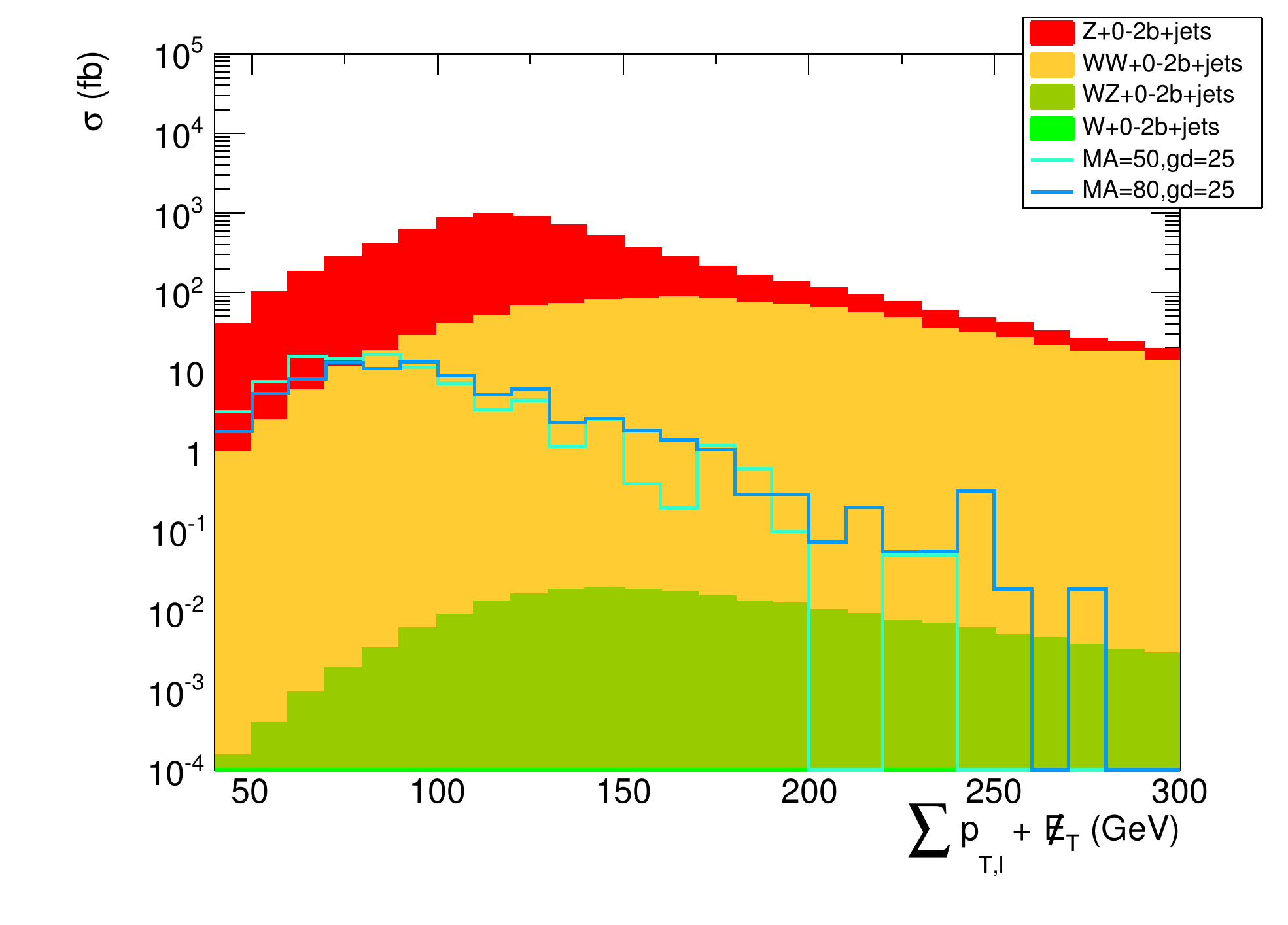}
}
\caption{Kinematic distributions of the signal and background for pseudoscalar mediated dilepton production, including backgrounds, with a $\mu^+\mu^-$ signature, with no cuts made. Upper left figure is the total scalar sum of visible transverse momentum, upper right figure is the leading lepton $p_T$, and lower figure is the scalar sum of the lepton $p_T$ and the $\met$. Distributions show distinctive differences in the distributions between signal and background.\label{fig:mumu}}
\end{figure*}

To illustrate how Seer was used for this project, the contents of the settings files are listed below. The contents of the {\file seer\_files.txt} file was:
\begin{verbatim}
## Single W plus 0-2 bjets plus light jets
W+0-2b+jets d 1 ./lhcofiles/bkgs/wmbb/run_01.lhco 5.81426 41257
...
W+0-2b+jets d 1 ./lhcofiles/bkgs/wpbb/run_01.lhco 9.76014 40566
...
## W+Z plus 0-2 bjets
WZ+0-2b+jets d 0 ./lhcofiles/bkgs/wmzbb/run_01.lhco 0.000426996 50000
...
WZ+0-2b+jets d 0 ./lhcofiles/bkgs/wpzbb/run_01.lhco 0.000693746 50000
...
## WW plus 0-2 bjets, includes ttbar production
WW+0-2b+jets d 0 ./lhcofiles/bkgs/wwbb/run_01.lhco 3.8891 50000
...
## gamma*/Z plus 0-2 bjets plus jets
Z+0-2b+jets d 0 ./lhcofiles/bkgs/llbb/run_01.lhco 1.35258 32926
...
## Signal files.
MA=80,gd=25 d 0 ./lhcofiles/signal/set_80_5_1.lhco 0.923511 50000
...
\end{verbatim}
The ellipses refer to multiple entries of similar files. For the backgrounds, events were broken down into subsets of diagrams, which necessitates overriding the automatic event weight calculator in Seer and adding the cross section (in pb) and number of events at the end of each line (note the use of the \texttt{d} flag when overriding the automatic event weight calculator). Backgrounds involving a single $W$ decaying to leptons do not meet the dilepton criteria, and so a light jet must fake either an electron or a tau lepton to contribute to the signal. This was calculated by using Seer to estimate fakes instead of Delphes. 

Dimuon events do not suffer from large fake lepton backgrounds, but $b$-jet faking was a significant issue. For the $e\mu$ analysis, however, jets were considered to fake electrons, and so the single $W$+jets channel could produce events that faked the signal. In this case, the flag of \texttt{1} for the single $W$ backgrounds shown above corresponded to scheme 1 of {\file seer\_fakes.txt}, which contained:
\begin{verbatim}
Rates
jet ele 0.0001
jet tau 0.001
jet bjt 0.001
Schema
scheme 0
scheme 1
addreal 1 1 muo, pt 10 99999, eta -2.5 2.5
addreal 1 2 bjt, pt 10 99999, eta -2.5 2.5
addfake jet ele, pt 10 99999, eta -2.5 2.5
scheme 2
...
\end{verbatim}
This is the entry for the $e\mu$ analysis, where the muon is expected from the decay of the $W$, and the analysis requires that there is at least 1 $b$-jet. For events without any $b$-jets, a second entry of \texttt{addfake jet bjt, pt 10 99999, eta -2.5 2.5} can be added to the same scheme. For each event that meets that criteria, these settings instruct Seer to randomly select a jet with $p_T > 10$~GeV and $|\eta| < 2.5$ and switch its type to an electron. The fake rate for this is $\not\!\epsilon_j^e = 0.0001$, and this factor is then applied to all such events. Not included here for brevity is a similar inclusion of dilepton events with zero $b$-jets, in which a light jet fakes a $b$-jet.

The {\file seer\_cuts.txt} file for the dimuon analysis was as follows:
\begin{verbatim}
## New Cut Card - double number signs at the start of the line indicate a comment
## Note: B-jets and hadronic taus are also considered jets unless they are explicitly tagged/removed!
Signal ## Do not alter this line!
2 2 muo, pt 10 99999, eta -2.5 2.5
1 2 bjt, pt 10 99999, eta -2.5 2.5
Extra ## Do not alter this line!
##THIS IS CMS TRIGGER
trigger ele pt 35 0 0 0
trigger ele pt 23 12 0 0
trigger muo pt 25 0 0 0
trigger muo pt 17 10 0 0
trigger gam pt 80 0 0 0
trigger gam pt 40 25 0 0
trigger tau pt 59 59 0 0
trigger jet pt 657 0 0 0
trigger jet pt 247 247 247 0
trigger jet pt 113 113 113 113
trigger tae pt 45 19 0 0
trigger tam pt 40 15 0 0
trigger jat pt 180 123 0 0
trigger tat pt 86 65 0 0
trigger bjt pt 237 0 0 0
trigger eam pt 23 10 0 0
trigger eam pt 12 23 0 0
##IGNORES
ignore ele with pt between 0 10
ignore ele with eta over 2.47
ignore muo with pt between 0 10
ignore muo with eta over 2.4
ignore gam with pt between 0 10
ignore gam with eta over 2.5
ignore jet with pt between 0 40
ignore jet with eta over 2.5
ignore bjt with pt between 0 10
ignore bjt with eta over 2.5
##CUTS
addcut all with osl over 0
##Below are the cuts for the 2u case
addcut lep with pt over 50
addcut all with ht over 120
addcut all with htl over 120
\end{verbatim}
The signal tagging required exactly two muons and either one or two $b$-jets. An estimate of the CMS trigger thresholds was included, and detector limitations for electrons, muons and jets were implemented in the \texttt{ignore} section. 

The cuts included in this list are: an opposite sign lepton cut (eliminating same-sign lepton pairs); a $p_T<50$~GeV on the leading lepton cut (events with $p_T>50$~GeV were removed); a total scalar sum of $p_T$ of visible objects of $H_T<120$~GeV cut; and a scalar sum of the missing transverse momentum and $p_T$ of the leptons cut of $\not\!\!H_T^\ell < 120$.

The {\file seer\_plots.txt} file for Figure \ref{fig:mumufin} included the settings:
\begin{verbatim}
PlotType = 35
PlotType2 = 0
PlotMaxX = 150
PlotMinX = 0
PlotMaxY = 10000
PlotMinY = 0.0001
NumBinX = 150
NumBinY = 30
ChooseLn = 1
Normalize = 0
PlotTxt = 1u1u1
SigPref = MA
FileType = pdf
\end{verbatim}
Setting \texttt{PlotType2} to 0 creates a single variable histogram, and the \texttt{NumBinX} value sets the bin width to be 1~GeV. Based on the \texttt{SigPref} entry, all entries in {\file seer\_files.txt} that have a tag starting with ``MA" are treated as signal files and plotted as outlines, while all others are treated as backgrounds. Based on the \texttt{FileType} entry, the histogram was output as a portable document format (pdf) file.

\begin{figure*}[tb]
\centering
\mbox{\includegraphics[width=0.45\textwidth,clip]{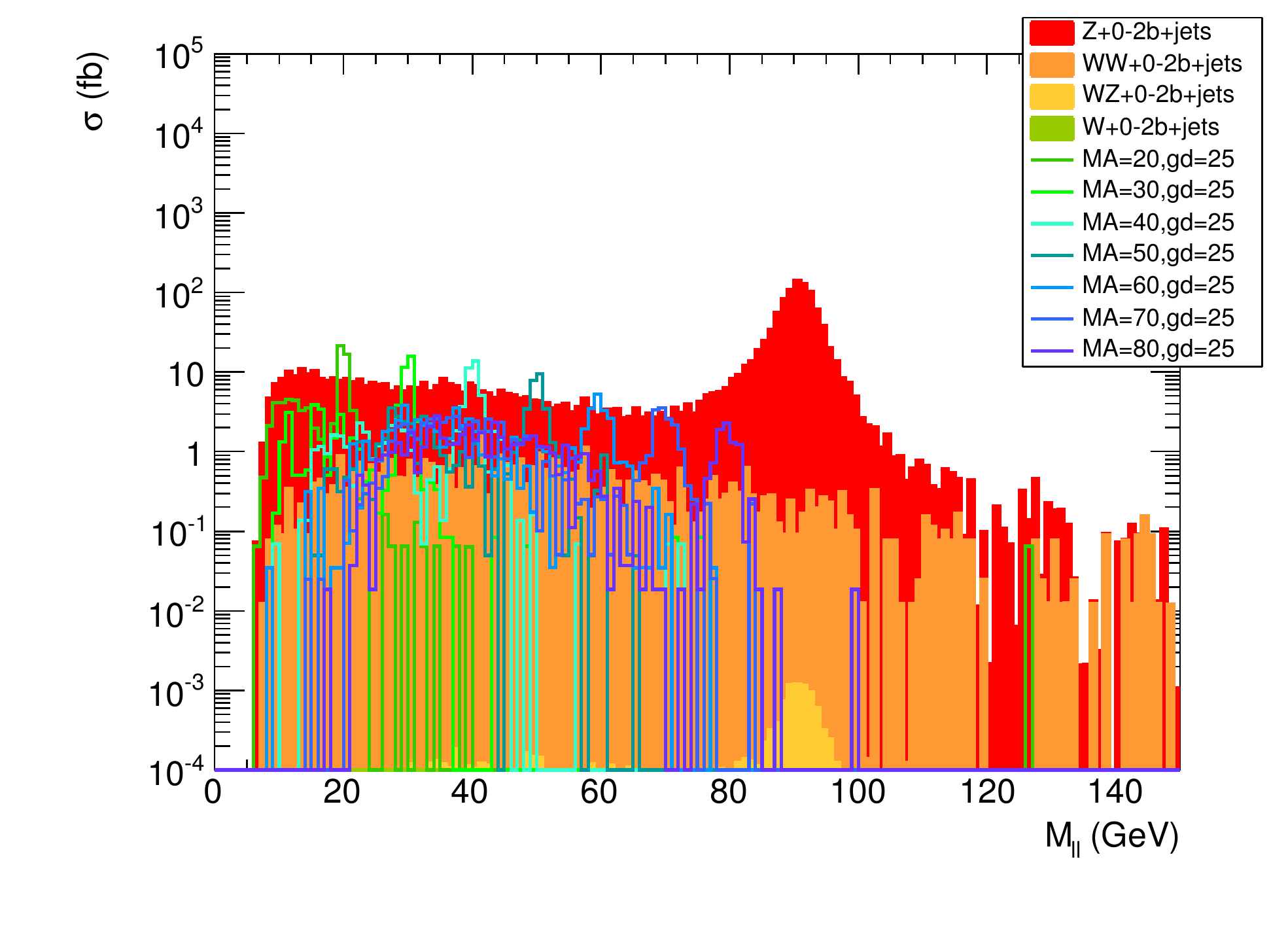}}
\caption{Invariant mass distribution of pseudoscalar mediated $\mu^+\mu^-$ production, as well as SM backgrounds, with cuts made to enhance signal over the background.\label{fig:mumufin}}
\end{figure*}

With the small bin sizes in Figure \ref{fig:mumufin}, the event rates per bin show a significant statistical fluctuation in bins where few backgrounds passed the tagging and cut requirements. This motivated the need for generating 40.7 million events, in order to get an accurate estimate of the backgrounds in each bin. For the ROOT file format, the file size for 50k events is larger than 500~MB. Performing this same analysis using information stored in ROOT files would require more than 400~GB of disk space. The time taken for the generation of these plots was approximately 20 minutes (MacBook Pro 2.66GHz i7, 8GB 1066 DDR3 RAM, 5400 RPM HD).

\begin{figure*}[tb]
\centering
\mbox{
\includegraphics[width=0.45\textwidth]{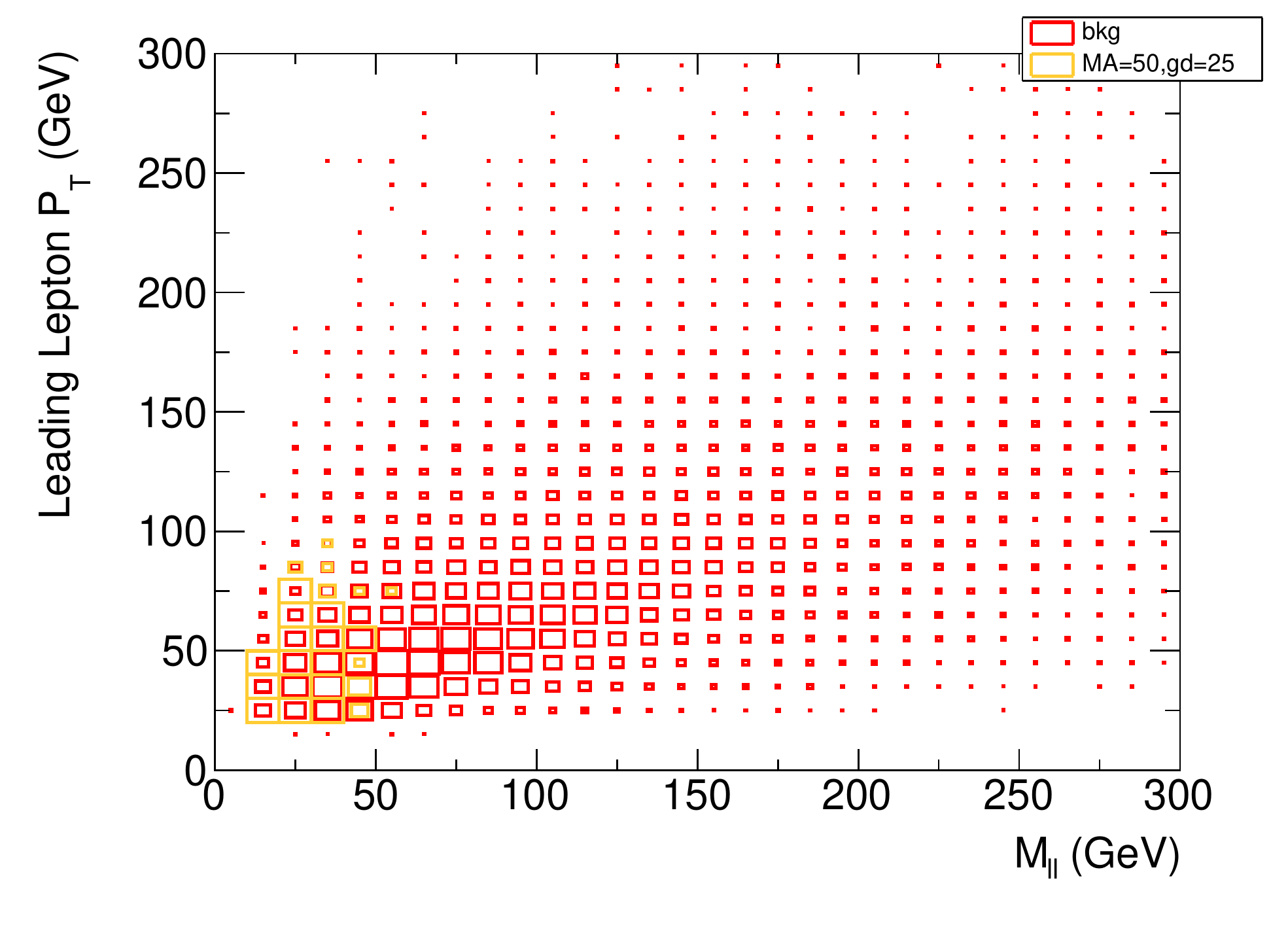}}\\
\mbox{
\includegraphics[width=0.45\textwidth]{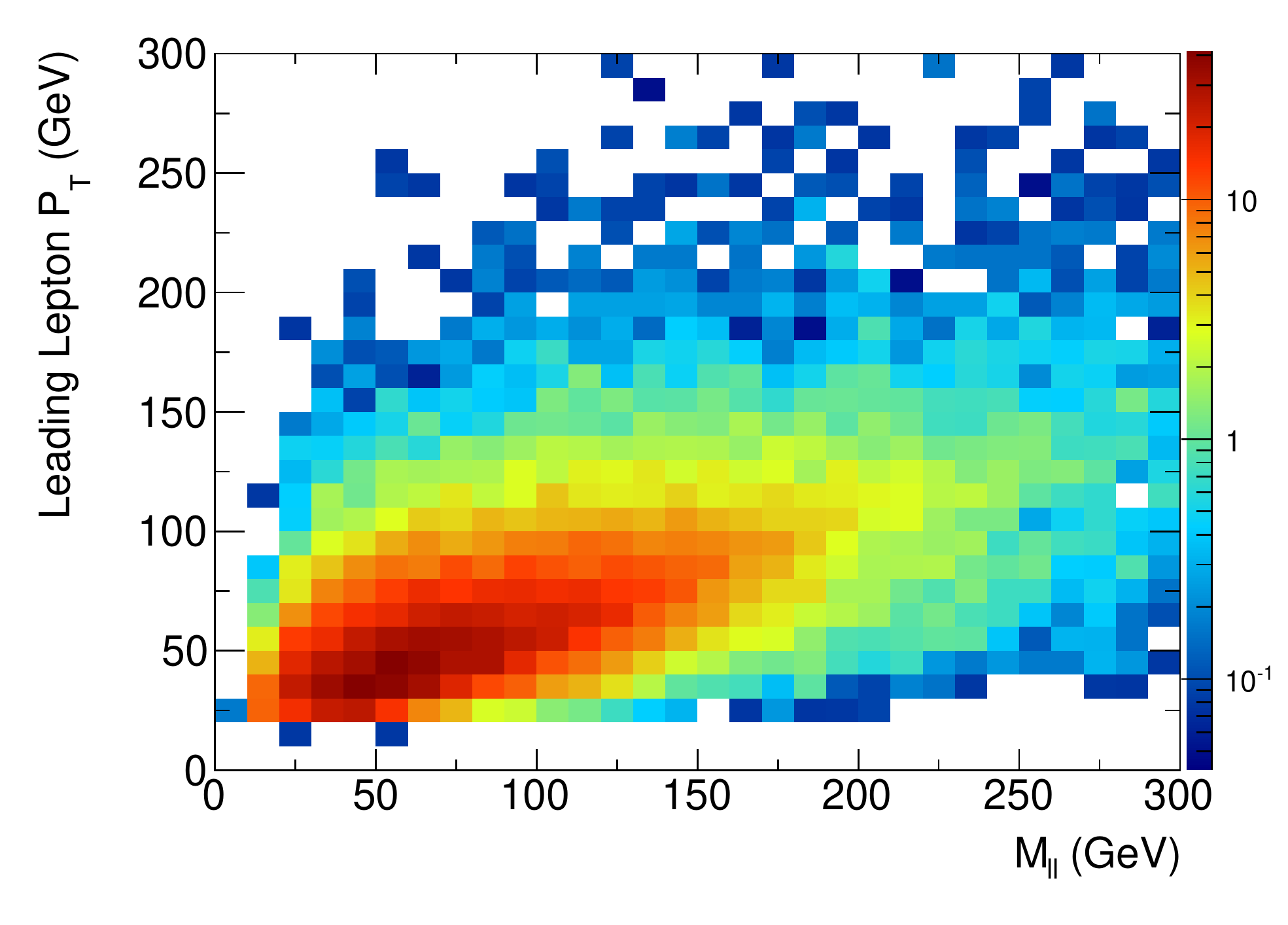}
\includegraphics[width=0.45\textwidth]{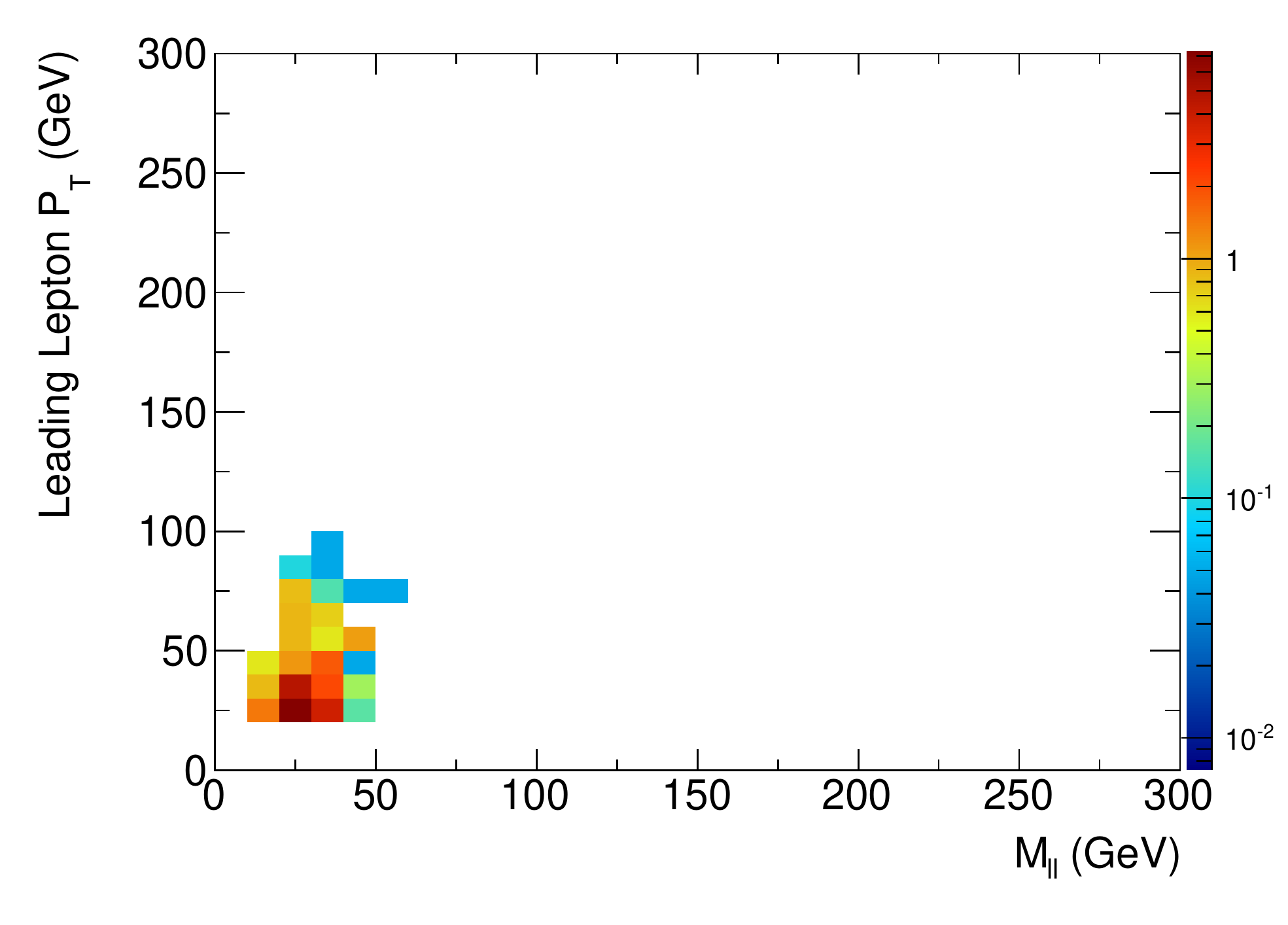}}
\caption{Examples of the types of 2D histograms that can be generated by Seer. The box histogram (top) is produced when multiple tags are included in the {\file seer\_files.txt} list. Box sizes represent the relative bin value. The coloured spectral histograms (lower) are produced when only a single tag is listed in the {\file seer\_files.txt} list. Without setting Seer to produce normalized cross sections, the $z$-axis values are $\sigma$ values measured in fb.\label{fig:2d}}
\end{figure*}

Seer also allows for alternative ways of visualizing the events. The plots in Figure \ref{fig:2d} were produced by making a few changes to the settings files in Seer, but using the same input files. The signal is switched to the $e+\mu$ channel, which does not have the large invariant mass peak, by adjusting the appropriate signal lines in the {\file seer\_cuts.txt} file, shown below:
\begin{verbatim}
Signal ## Do not alter this line!
1 1 ele, pt 10 99999, eta -2.5 2.5
1 1 muo, pt 10 99999, eta -2.5 2.5
1 2 bjt, pt 10 99999, eta -2.5 2.5
\end{verbatim}
Using this tagging, Figure \ref{fig:2d} was produced by combining all four background types into a single tag (bkg), and setting {\file seer\_plots.txt} to:
\begin{verbatim}
PlotType = 35
PlotType2 = 5
PlotMaxX = 300
PlotMinX = 0
PlotMaxY = 300
PlotMinY = 0
NumBinX = 30
NumBinY = 30
ChooseLn = 1
Normalize = 0
PlotTxt = 1e1u
SigPref = MA
FileType = pdf
\end{verbatim}
Additionally, all of the backgrounds were given an identical tag in {\file seer\_files.txt}, and only one signal set was included. The top of Figure \ref{fig:2d} was produced with both the signal and background entries in {\file seer\_files.txt}, while the lower two were produced by commenting out the signal or background, respectively. This visual representation of the phase space is useful for identifying cut regions.

\subsection{Reproducing Experimental Analyses}

Seer can also be used to test MC signal data against existing experimental results. As with any such endeavour, the precise details of the detector simulation (such as in Delphes) and method for accounting for NLO effects can have a significant impact on the results. Included in this section are  the results of using Seer to reproduce three experimental SUSY searches: 2-6 jet inclusive search \cite{TheATLAScollaboration:2013fha}, tri-lepton search \cite{ATLAS:2013rla}, di-lepton search \cite{TheATLAScollaboration:2013hha}. The routines to perform the cuts for each of these studies are implemented in Seer already in {\file calculations.cpp}, allowing users to perform these same searches. In each case, the details of the Delphes settings used are not included, as this is outside the scope of Seer.

Another challenge in reproducing experimental searches is in accounting for combined exclusion limits. The experimental groups typically use the $CL_s$ method for determining exclusion constraints, which accounts for signal excesses from multiple signal regions, and accounts for correlations between systematic uncertainties present. The correlation matrix for the systematic uncertainties are not currently provided by any of the experimental groups, and so theorists are unable to perform this same analysis. In the studies reproduced here, exclusion contours were generated by assuming a boolean addition of the exclusions from each independent signal region. This naturally results in a more conservative contour, leading to a natural negative bias, especially near regions that are excluded by multiple signal regions simultaneously.

\subsubsection{ATLAS 2-6 jet inclusive search}

The 2-6 jet inclusive search \cite{TheATLAScollaboration:2013fha} that was reproduced involved the mSUGRA scan over $m_{1/2}$ vs $m_0$, with $\tan\beta = 30$, $A_0 = -2m_0$ and $\mu>0$. Contrary to the conference note, which used Herwig++ to generate events, this reproduction employed MG5+aMC@NLO to generate 50k events of the fully inclusive di-sparticle cross section (all possible pairings of sparticles) at leading order, without any additional hard element jets. These states were then decayed and hadronized using the standard Pythia 6.4 implementation with MG5+aMC@NLO, and Delphes 3.0 was employed to simulate detector effects. The parameters $m_0$ and $m_{1/2}$ were scanned over the values $m_0 \sim 700-5700$~GeV (step size 500 GeV) and $m_{1/2} \sim 300-900$~GeV (step size 100 GeV).

The search was reproduced using the following {\file seer\_cuts.txt} settings:
\begin{verbatim}
Signal ## Do not alter this line!
2 99 jet, pt 10 99999, eta -5 5
Extra ## Do not alter this line!
trigger jat pt 80 100 0 0
##IGNORES
ignore ele with pt between 0 10
ignore ele with eta over 2.47
ignore muo with pt between 0 10
ignore muo with eta over 2.4
ignore gam with pt between 0 10
ignore gam with eta over 2.5
ignore bjt with pt between 0 20
ignore bjt with eta over 2.5
ignore ljt with pt between 0 20
ignore ljt with eta over 4.5
##CUTS
## A-loose
addcut all with jetA over 1
## A-medium
## addcut with jetA over 2
## B-medium
## addcut with jetB over 1
## B-tight
## addcut with jetB over 2
## C-medium
## addcut with jetC over 1
## C-tight
## addcut with jetC over 2
## D
## addcut with jetD over 1
## E-loose
## addcut with jetE over 1
## E-medium
## addcut with jetE over 2
## E-tight
## addcut with jetE over 3
\end{verbatim}
where the appropriate signal region (A through E) was uncommented separately for each run of Seer. All LHCO files for each data point were included in {\file seer\_files.txt}, resulting in a total of 10 runs of Seer, once for each signal region. Due to the large number of LHCO used, the plotting features were disabled in {\file seer\_plots.txt}, as distributions of events were irrelevant and only the overall summary provided in the text file output was needed. A script, {\file analysis.cpp} (included in the {\file Results} folder of Seer) was used to read through the output files and generate a Mathematica script to plot the exclusion contour.  

The results of the calculations using Seer are shown in Figure \ref{fig:ex_047}. The leading order cross sections agree with the values in the conference note to within about 20\%, depending on the point in parameter space, but are systematically lower. Due to the large number of QCD events in the fully inclusive cross section, the effect of ISR jets and other NLO effects can be quite sizeable. Instead of incorporating them, these effects were accounted for using a K-factor of 35\%. \footnote{This value is chosen based on the K-factors listed on \url{https://twiki.cern.ch/twiki/bin/view/Sandbox/TestTopic232} for the mSUGRA benchmark points. Ideally, the K-factor will vary based on the particular parameter points. A flat K-factor was chosen as a sufficient approximation to vet the Seer code.} The excellent agreement between Seer and the experimental results suggests that Seer is sufficiently capable of analyzing and reproducing the experimental study.

\begin{figure*}[tb]
\centering
\mbox{\includegraphics[width=0.45\textwidth,clip]{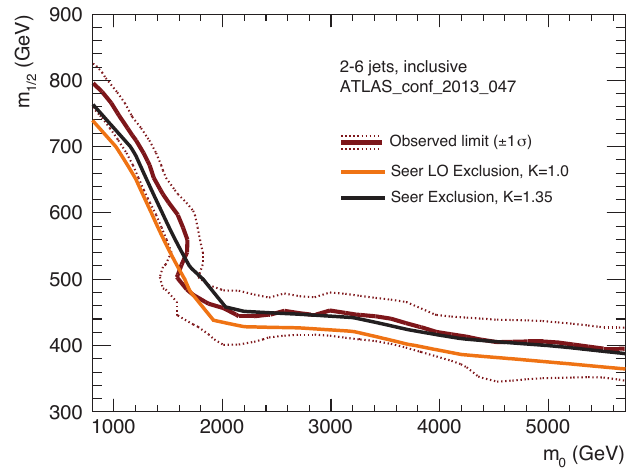}}
\caption{Exclusion plot for inclusive jet+$\not\!\!{E}_T$ production in mSUGRA from Seer (orange at LO, black with K-factor), as compared to the observed limit from the originating ATLAS study (red) \cite{TheATLAScollaboration:2013fha}.\label{fig:ex_047}}
\end{figure*}

\subsubsection{ATLAS tri-lepton search}

The tri-lepton search that was reproduced \cite{ATLAS:2013rla} focused on a simplified version of the pMSSM, where the squarks, gluons and Higgsinos are decoupled. This leaves three remaining relevant parameters, $M_1 \sim M_{\tilde{\chi}_1^0}$ and $M_2 \sim M_{\tilde{\chi}_2^0} \sim M_{\tilde{\chi}_1^\pm}$ which are varied, and a fixed $\tan\beta = 10$. The process $pp \rightarrow \tilde{\chi}_1^\pm \tilde{\chi}_2^0$ was generated with 50k events, and the same process of production, decay, hadronization and detector simulation was performed as discussed in the previous section.

The {\file seer\_cuts.txt} settings used to reproduce this study were:
\begin{verbatim}
Signal ## Do not alter this line!
3 99 lep, pt 10 99999, eta -2.5 2.5
0 99 jet, pt 20 99999, eta -2.5 2.5
0 99 gam, pt 10 99999, eta -2.5 2.5
Extra ## Do not alter this line!
##THIS IS ATLAS TRIGGER
trigger ele pt 25 0 0 0
trigger ele pt 15 15 0 0
trigger muo pt 20 0 0 0
trigger muo pt 10 10 0 0
trigger gam pt 60 0 0 0
trigger gam pt 20 20 0 0
trigger jet pt 400 0 0 0
trigger jet pt 165 165 165 0
trigger jet pt 110 110 110 110
trigger jat pt 70 70 0 0
trigger tat pt 35 45 0 0
trigger eam pt 20 10 0 0
trigger eam pt 10 20 0 0
##IGNORES
ignore ele with pt between 0 10
ignore ele with eta over 2.47
ignore muo with pt between 0 10
ignore muo with eta over 2.4
ignore gam with pt between 0 10
ignore gam with eta over 2.47
ignore jet with pt between 0 20
ignore jet with eta over 2.5
addcut bjt with num over 0
addcut all with a3lp over 1
\end{verbatim}
where the numerical value of the \texttt{addcut all with a3lp over 1} line was varied from 1-6 for each signal region. All other aspects of the analysis were similar as to the 2-6 jets study discussed previously. The results of this are shown in Figure \ref{fig:ex_035}.

In this case, the exclusion region found using Seer is in excellent agreement for $m_{\tilde{\chi}_2^0} - m_{\tilde{\chi}_1^0} \gg m_Z$. In the region where the spectrum compresses, multiple signal regions provide exclusions. This results in an enhanced exclusion using the full power of the $CL_s$ method, as discussed, but an inability to precisely reproduce the exclusions using Seer. This is a limitation that cannot be resolved currently without more information regarding the correlations of systematic uncertainties in the experimental signal regions. 

The original study employed the NLO cross sections from Prospino 2.0 \cite{Beenakker:1996ed}, however the reproduction in Seer uses LO cross sections. Appropriate K-factors in this region are approximately 10-20\%, based on a randomized parameter point check in Prospino over the parameter space explored. An overall 15\% K-factor has been included as a separate contour in Figure \ref{fig:ex_035}. Regions where the K-factor significantly increases the exclusion typically have high acceptance rates. Alternatively, where the K-factor exclusion line is close to the LO exclusion, the acceptance rates are low.

\begin{figure*}[tb]
\centering
\mbox{\includegraphics[width=0.45\textwidth,clip]{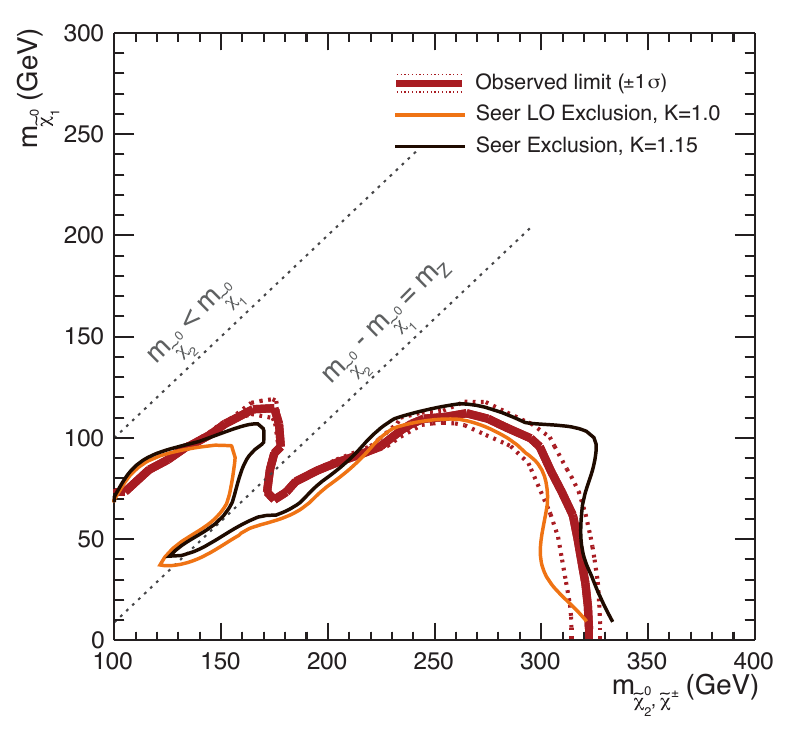}}
\caption{Exclusion plot for chargino+neutralino production in the three lepton final state from Seer (orange at LO, black with K-factor), as compared to the observed limit from the originating ATLAS study (red) \cite{ATLAS:2013rla}. Lightest neutralino masses smaller than 10 GeV were not generated as part of this analysis.\label{fig:ex_035}}
\end{figure*}

\subsubsection{ATLAS di-lepton search}

The di-lepton search that was reproduced \cite{ATLAS:2013rla} focused on a simplified version of the pMSSM, where right handed sleptons (degenerate $\tilde{e}_R$ and $\tilde{\mu}_R$) decay directly to a lepton and a bino-type lightest neutralino. This results in a similar three parameter subset of the pMSSM, with a scan over the mass of the sleptons $m_{\tilde{\ell}}$ and the bino mass $M_1 \sim m_{\tilde{\chi}_1^0}$, and a fixed $\tan\beta=10$. The process $pp \rightarrow \tilde{\ell}_R^+ \tilde{\ell}_R^-$ was generated with 50k events using the same generation, decay, hadronization and detector simulation discussed previously.

The {\file seer\_cuts.txt} settings used to reproduce this study were:
\begin{verbatim}
Signal ## Do not alter this line!
2 2 lep, pt 10 99999, eta -2.5 2.5
Extra ## Do not alter this line!
##THIS IS ATLAS TRIGGER
trigger ele pt 25 0 0 0
trigger ele pt 15 15 0 0
trigger muo pt 20 0 0 0
trigger muo pt 10 10 0 0
trigger gam pt 60 0 0 0
trigger gam pt 20 20 0 0
trigger jet pt 400 0 0 0
trigger jet pt 165 165 165 0
trigger jet pt 110 110 110 110
trigger jat pt 70 70 0 0
trigger tat pt 35 45 0 0
trigger eam pt 20 10 0 0
trigger eam pt 10 20 0 0
##IGNORES
ignore ele with eta over 2.47
ignore muo with eta over 2.4
ignore gam with pt between 0 20
ignore gam with eta over 2.4
ignore jet with pt between 0 20
ignore jet with eta over 4.5
addcut all with a2lp over 2
\end{verbatim}
where the numerical value of the \texttt{addcut all with a2lp over 1} line was varied from 1-5 for each signal region. All other aspects of the analysis were similar as to the 2-6 jets study discussed previously.

Very similar exclusion regions were found, however for the region of large $m_{\tilde{\ell}}$ the Seer exclusion does not lie within the $\pm 1 \sigma$ boundary for the exclusion. This is due to the very low uncertainties in the signal, which results from the exclusive nature of the search (all events are vetoed for the presence of any jet with $p_T>20$~GeV). One particular explanation for the difference in the exclusions is that Delphes employs a flat efficiency for lepton identification. This is one area where CheckMATE has improved upon the standard Delphes implementation available with the MG5+aMC@NLO package by incorporating many more details regarding lepton tagging and identification. None-the-less, Seer performs exceptionally well in reproducing this study, even with the standard Delphes treatment of leptons.

A similar K-factor to the tri-lepton study was included in this analysis, to show the dependence on the signal cross section. Appropriate K-factors found using Prospino are on the order of 10-20\%, however the jet veto employed in this analysis would suggest that an appropriate K-factor is more challenging to determine. Additionally, and since the K-factor is parameter point dependent, this value is included purely as a qualitative descriptor of the effect of increasing the signal cross section on the exclusion results.

\begin{figure*}[tb]
\centering
\mbox{\includegraphics[width=0.45\textwidth,clip]{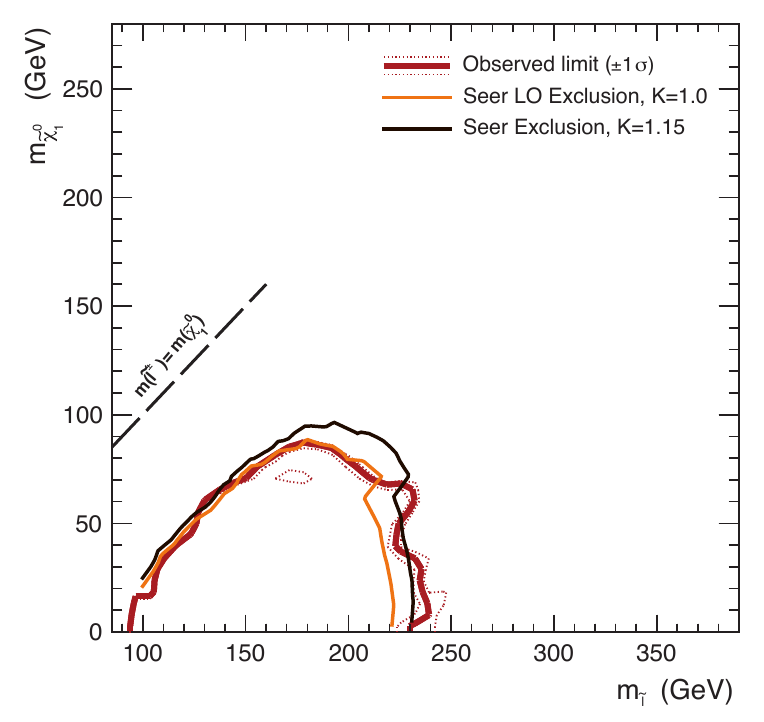}}
\caption{Exclusion plot for right handed slepton pair production from Seer (orange at LO, black with K-factor), as compared to the observed limit from the originating ATLAS study (red) \cite{TheATLAScollaboration:2013hha}. Slepton masses smaller than 100 GeV were not generated as part of this analysis.\label{fig:ex_049}}
\end{figure*}

\section{Summary}

Seer is a useful tool for analyzing LHCO output files for collider processes that have gone through a detector simulation like PGS or Delphes. With a simple structure to use, and an easy-to-modify script system for the implementation of new cuts and kinematic variables, Seer can be rapidly employed for exploring kinematic distributions as well as an analysis of trigger response, cut efficiencies, and even full analyses of experimental processes.

There are a number of primary uses for this tool. The most obvious would be in visualizing kinematic distributions and exploring the kinematics of a signal. The second would be in reproducing experimental analyses. The third is in testing new kinematic variables and distributions. For example, stransverse variables are of particular interest for SUSY searches, and new variants are still being developed. The Seer code structure provides a simple framework that allows for rapid implementation of these new test variables.

Alternative code packages such as MadAnalysis 5 and CheckMATE include different features and different interfaces, with different learning curves. Seer presents an easily usable, easily modifiable, powerful and effective way to analyze the results of simulation data.

\section*{Acknowledgements}
The author would like to thank Alejandro de la Puente and Jonathan Kozaczuk for encouragement to share this package, and Carlos Wagnar's tongue-in-cheek encouragement with his statement, ``There is no point in naming it if you aren't going to release it," or something to that effect. Additionally, the author would like to thank David Morrissey for his patience and encouragement while Seer was being written, and J.P. Archambault for contributions to the style code for ROOT. This work was supported in part by the National Science and Engineering Research Council of Canada~(NSERC).


\begin{thebibliography}{99}
\bibitem{Alwall:2014hca} 
  J.~Alwall, R.~Frederix, S.~Frixione, V.~Hirschi, F.~Maltoni, O.~Mattelaer, H.-S.~Shao and T.~Stelzer {\it et al.},
  JHEP {\bf 1407}, 079 (2014)
  [arXiv:1405.0301 [hep-ph]].

\bibitem{Sjostrand:2007gs} 
  T.~Sjostrand, S.~Mrenna and P.~Z.~Skands,
  Comput.\ Phys.\ Commun.\  {\bf 178}, 852 (2008)
  [arXiv:0710.3820 [hep-ph]].
  
\bibitem{Bahr:2008pv} 
  M.~Bahr, S.~Gieseke, M.~A.~Gigg, D.~Grellscheid, K.~Hamilton, O.~Latunde-Dada, S.~Platzer and P.~Richardson {\it et al.},
  Eur.\ Phys.\ J.\ C {\bf 58}, 639 (2008)
  [arXiv:0803.0883 [hep-ph]].
  
\bibitem{Gleisberg:2003xi} 
  T.~Gleisberg, S.~Hoeche, F.~Krauss, A.~Schalicke, S.~Schumann and J.~C.~Winter,
  JHEP {\bf 0402}, 056 (2004)
  [hep-ph/0311263].
  
\bibitem{Reuter:2014ema} 
  J.~Reuter, F.~Bach, B.~Chokoufe, W.~Kilian, T.~Ohl, M.~Sekulla and C.~Weiss,
  arXiv:1410.4505 [hep-ph].
  
\bibitem{Carena:2000yx} 
  M.~S.~Carena {\it et al.}  [Higgs Working Group Collaboration],
  hep-ph/0010338.
  
\bibitem{Selvaggi:2014mya} 
  M.~Selvaggi,
  J.\ Phys.\ Conf.\ Ser.\  {\bf 523}, 012033 (2014).
  
\bibitem{Conte:2014xya} 
  E.~Conte, B.~Dumont, B.~Fuks and T.~Schmitt,
  arXiv:1410.2785 [hep-ph].
  
\bibitem{Drees:2013wra} 
  M.~Drees, H.~Dreiner, D.~Schmeier, J.~Tattersall and J.~S.~Kim,
  Comput.\ Phys.\ Commun.\  {\bf 187}, 227 (2014)
  [arXiv:1312.2591 [hep-ph]].
  
\bibitem{Alwall:2006yp} 
  J.~Alwall, A.~Ballestrero, P.~Bartalini, S.~Belov, E.~Boos, A.~Buckley, J.~M.~Butterworth and L.~Dudko {\it et al.},
  Comput.\ Phys.\ Commun.\  {\bf 176}, 300 (2007)
  [hep-ph/0609017].
  
\bibitem{Antcheva:2009zz} 
  I.~Antcheva, M.~Ballintijn, B.~Bellenot, M.~Biskup, R.~Brun, N.~Buncic, P.~Canal and D.~Casadei {\it et al.},
  Comput.\ Phys.\ Commun.\  {\bf 180}, 2499 (2009).
  
\bibitem{Walker:2012vf} 
  J.~W.~Walker,
  arXiv:1207.3383 [hep-ph].
  
\bibitem{Rogan:2010kb} 
  C.~Rogan,
  arXiv:1006.2727 [hep-ph].
  
\bibitem{Chatrchyan:2012uea} 
  S.~Chatrchyan {\it et al.}  [CMS Collaboration],
  Phys.\ Rev.\ Lett.\  {\bf 111}, no. 8, 081802 (2013)
  [arXiv:1212.6961 [hep-ex]].
  
\bibitem{Cheng:2008hk} 
  H.~C.~Cheng and Z.~Han,
  JHEP {\bf 0812}, 063 (2008)
  [arXiv:0810.5178 [hep-ph]].
  
\bibitem{Lester:1999tx} 
  C.~G.~Lester and D.~J.~Summers,
  Phys.\ Lett.\ B {\bf 463}, 99 (1999)
  [hep-ph/9906349].
  
\bibitem{Barr:2003rg} 
  A.~Barr, C.~Lester and P.~Stephens,
  J.\ Phys.\ G {\bf 29}, 2343 (2003)
  [hep-ph/0304226].
  
\bibitem{ATL-PHYS-PUB-2013-004}
  ATLAS Collaboration,
  ATL-PHYS-PUB-2013-004.
    
\bibitem{Kozaczuk:2015bea} 
  J.~Kozaczuk and T.~A.~W.~Martin,
  arXiv:1501.07275 [hep-ph].
  
\bibitem{TheATLAScollaboration:2013fha} 
  The ATLAS collaboration,
  ATLAS-CONF-2013-047, ATLAS-COM-CONF-2013-049.
  
\bibitem{ATLAS:2013rla} 
  [ATLAS Collaboration],
  ATLAS-CONF-2013-035, ATLAS-COM-CONF-2013-042.
  
\bibitem{TheATLAScollaboration:2013hha} 
  The ATLAS collaboration,
  ATLAS-CONF-2013-049, ATLAS-COM-CONF-2013-050.
  
\bibitem{Beenakker:1996ed} 
  W.~Beenakker, R.~Hopker and M.~Spira,
  hep-ph/9611232.
  
  
\end{thebibliography}
\end{document}